\documentclass[fleqn,usenatbib]{mnras}

\usepackage{newtxtext,newtxmath}
\usepackage{mathptmx}
\usepackage{txfonts}

\usepackage[T1]{fontenc}

\DeclareRobustCommand{\VAN}[3]{#2}
\let\VANthebibliography\thebibliography
\def\thebibliography{\DeclareRobustCommand{\VAN}[3]{##3}\VANthebibliography}

\usepackage{graphicx}	
\usepackage{amsmath}	
\usepackage{amssymb}	
\newcommand\apspr{Astrophys.~Space~Phys.~Res.} 
\usepackage[export]{adjustbox}

\title{The Transient Ultra-luminous X-ray Source, ULX-4, in M51}

\author[S. Allak et al.]{S. Allak,$^{1,2}$\thanks{E-mail:0417allaksinan@gmail.com} A. Akyuz,$^{3,2}$ 
İ. Akkaya Oralhan,$^{4}$
S. Avdan,$^{2}$
N. Aksaker,$^{5,2}$ 
A. Vinokurov,$^{6}$ \and
F. Soydugan,$^{1,7}$
E. Sonbas,$^{8,9}$ 
and K. S. Dhuga$^{9}$
\\ \\
$^1$Department of Physics, University of Çanakkale Onsekiz Mart, 17100, Çanakkale, Turkey\\
${^2}$Space Science and Solar Energy Research and Application Center (UZAYMER), University of Çukurova, 01330, Adana, Turkey\\
$^3$Department of Physics, University of Çukurova, 01330, Adana, Turkey\\
$^4$Department of Astronomy and Space Sciences, Erciyes University, 38039, Kayseri, Turkey\\
$^5$Adana Organised Industrial Zones Vocational School of Technical Science, University of Çukurova, 01410, Adana, Turkey\\
$^6$Special Astrophysical Observatory of the Russian AS, Nizhnij Arkhyz, Russia\\
$^7$Astrophysics Research Centre and Ulupınar Observatory, University of Çanakkale Onsekiz Mart, 17100, Çanakkale, Turkey\\
$^8$Adiyaman University, Department of Physics, 02040 Adiyaman, Turkey\\
$^9$Department of Physics, The George Washington University, Washington, DC 20052, USA\\
\\
}
\date{Accepted 2021 December 14. Received 2021 December 2; in original form 2021 June 14.}

\pubyear{2021}

\begin{document}
\label{firstpage}
\pagerange{\pageref{firstpage}--\pageref{lastpage}}
\maketitle

\begin{abstract}

We present the results of a temporal and spectral analysis of the transient source ULX-4 in the galaxy M51. The data used were drawn from {\it Chandra}, {\it XMM-Newton} and {\it Swift-XRT} archives, spanning the years 2000-2019. The X-ray flux of the source is seen to vary by two orders of magnitudes within a month but a short-term variability was not observed over the time intervals of 100-2000 second in the 0.3$-$10 keV energy band. We find some evidence for the existence of bi-modality feature in the flux distribution of ULX-4.
We identified two optical sources as possible counterparts within an error radius of 0.$\arcsec$18 at 95$\%$ confidence level for ULX-4 based on the archival {\it HST}/ACS and {\it HST}/WFC3 data. Blackbody fits of the spectral energy distributions (SEDs) indicate the spectral type to be B-type stars. One of these counterparts exhibits a low-amplitude optical periodicity of 264 $\pm$ 37 days in the F606W filter; if we assume this apparent periodicity is associated with the orbital motion of the donor, then it is more likely that the donor is a red supergiant satisfying the long periodicity and accretion via Roche-lobe overflow. Consequently, the SED would then have to be interpreted as a superposition of emissions from a cold donor and a hot flow component, most likely from an accretion disk. If, on the other hand, the periodicity is super orbital in nature i.e., due to possible interactions of the compact object with a circumstellar disk, the donor could then be a Be/X star hosting a neutron star.
\end{abstract}

\begin{keywords}
galaxies: individual (M51) X-rays: binaries–X-rays: general
\end{keywords}



\section{Introduction}

Ultraluminous X-ray Sources (ULXs) are point-like sources in external galaxies with an X-ray luminosity (L$_{X}$ $\geq$ 10$^{39}$ erg s$^{-1}$) exceeding the Eddington limit for a stellar-mass black hole. They are not located in the center of their host galaxies. This rules out the possibility of them being a supermassive black hole as a source of high emission (e.g. \citealp{2017ARA&A..55..303K, 2021AstBu..76....6F}, for a review). Recent studies tend to lean toward the majority of ULXs hosting stellar mass compact objects undergoing super-Eddington \citep{2017ApJ...834...77F,2021MNRAS.504..974G} rather than sub-Eddington accretions onto intermediate mass black holes (IMBHs) \citep{1999ApJ...519...89C, 2016A&A...595A.101T, 2017IJMPD..2630021M}.
In the case of super-Eddington accretion, the possible compact objects in ULXs may be either black holes (BHs) or neutron stars (NSs). While their relative numbers are still quite uncertain, the ULXs harboring NSs could be significantly more numerous than observed \citep{2017xru..conf..237W, 2020MNRAS.494.3611K}. So far, the presence of NSs in six ULXs has been confirmed by pulse detections and are known as pulsating ULXs (PULXs): M82 X2 \citep{2014Natur.514..202B}, NGC 7793 P13 \citep{2016ApJ...831L..14F,2017MNRAS.466L..48I},
NGC 5907 ULX-1 \citep{2017Sci...355..817I, 2017MNRAS.466L..48I}, NGC 300 ULX-1 \citep{2018MNRAS.476L..45C}, NGC 1313 X-2 \citep{2019MNRAS.488L..35S}, M51 ULX-7 \citep{2020ApJ...895...60R}.
In addition, \cite{2018NatAs...2..312B} discovered that M51 ULX-8 has a cyclotron resonance scattering feature. This feature is presumably caused by the interactions of charged particles with a strong magnetic field, thus suggesting that ULX-8 hosts a NS.
Also the first Galactic PULX (Swift J0243.6+6124) was discovered by the Swift/XRT telescope and X-ray pulsations at $\sim$ 9.86 s were detected in the 0.2-10 keV band \citep{2017ATel10809....1K}. Detected pulsations were then confirmed by data from different X-ray observatories \citep{2017GCN.21971....1B,2017ATel10866....1B,2018MNRAS.474.4432J,2021MNRAS.500..565B} and Fermi Gamma-Ray Burst Monitor (GBM) \citep{2017ATel10812....1J}.

It is known that ULXs are variable sources and their flux values in star-forming galaxies can vary by an order of magnitude \citep{2009ApJ...696.1712F}. However, a small fraction of ULXs for which flux variation reaches up to two orders of magnitude, are in the transient source class. Thanks to the new generation of X-ray observatories, the number of observations of these sources is rapidly increasing. This is likely to increase the discovery of new transient sources as well (e.g. \citealp{2012MNRAS.420.2969M,2013ApJ...772..126B,2018ApJ...864...64H,2019MNRAS.483.3566V,2020MNRAS.499.5682A, 2021MNRAS.501.1002W}). PULXs are sources observed in this class with their high flux variability reaching up to factors of 50 in flux variation (e.g. \citealp{2017ApJ...834...77F, 2020ApJ...890..166P}. This high variation has been interpreted in terms of the onset of a propeller mechanism \citep{2016MNRAS.457.1101T} and/or a super-orbital effect \citep{2019ApJ...873..115B}.

Investigation of ULXs in the optical wavelengths as well as the X-ray band can help reveal their physical nature. In particular, identification and characterization of the optical counterparts of ULXs plays an important role in advancing our understanding of the origin of the emission, whether it is primarily from the accretion disk and/or the donor star. Unfortunately, unique identification of the optical counterparts is not straightforward since most of them are located in very crowded regions or even in/near star clusters. Also, most optical counterparts are too faint (apparent magnitudes m$_{V}$ > 21 mag) to be detected by ground-based observations \citep{2011ApJ...737...81T,2013ApJS..206...14G,2018ApJ...854..176V}. However, there are few exceptions with m$_{V}$ $\sim$ 20 mag or even brighter, for example, NGC 7793 P13 \citep{2014Natur.514..198M}, NGC 300 ULX-1/SN2010da \citep{2016ApJ...830...11V} and UGC 6456 ULX \citep{2020ApJ...893L..28V}. The absolute magnitudes (M$_{V}$) of ULXs are in the range of $-$3 < M$_{V}$ < $-$8 mag \citep{2018ApJ...854..176V}. There are hundreds of ULXs but only about 20 have a single optical counterpart \citep{2011ApJ...737...81T,2013ApJS..206...14G,2019ApJ...875...68A,2020MNRAS.499.5682A}. Multiple optical counterparts have been detected for a number of ULXs \citep{2005MNRAS.356...12S,2007ApJ...658..999M,2019ApJ...875...68A,2019MNRAS.488.5935A, 2020MNRAS.499.2138A}.

The galaxy M51 (NGC 5194), also known as the Whirlpool galaxy, and its companion NGC 5195 are a pair of interacting galaxies located within the constellation Canes Venatici. The face-on spiral galaxy M51 has a low luminosity Seyfert 2 nucleus (at a distance of 9 Mpc; \cite{2014AJ....148..107R,2020MNRAS.491.1260S}). The remarkable aspect of the spiral arms of M51 is that they are very symmetric and serve as star formation factories as a result of the continuous and extremely regular dust lanes and the presence of diffuse interstellar medium \citep{1997ApJ...486L..95B}. As such, M51 is a good target for studying discrete X-ray populations, including ULXs. 

X-ray sources in M51 were investigated using the {\it ROSAT}/HRI observations. Among the nine X-ray sources detected, only three of them have luminosities of L$_{X}$ > 10$^{39}$ erg s$^{-1}$ \citep{2000MNRAS.315...98R, 2002ApJS..143...25C}.

Also 113 X-ray sources within the optical extents of the galaxy pair (NGC 5194/5195) were detected by \cite{2004ApJ...601..735T} using the 2000 and 2001 {\it Chandra} ACIS-S observations. Seven of these sources have $L_{\mathrm{X}}$ > $10^{39}$ erg s$^{-1}$ in the 0.5$-$8 keV range. Later, \cite{2005ApJ...635..198D} confirmed these seven ULXs and detected a new one using an {\it XMM-Newton} EPIC observation. They examined both short and long-term variability of the sources in detail.

\cite{2016ApJ...831...56U} presented the discovery of X-ray eclipses in two ULXs, ULX-1 and ULX-2 located in the same region of M51, by analyzing archival {\it Chandra} and {\it XMM-Newton} data. They reported the binary period of ULX-1 was constrained to be either $\simeq$ 6.3 days, or $\simeq$ 12.5$-$13 days. Also,
\cite{2018MNRAS.475.3561U} investigated the optical properties of these ULXs using {\it HST}, {\it LBT} and {\it VLA} observations and they determined that both ULXs were associated with ionized nebulae. \cite{2020ApJ...895...60R} discovered 2.8 s pulsations in the X-ray emission of the M51 ULX-7 using {\it XMM-Newton} data. The variable source ULX-7 has $L_{X}$ between $10^{39}$ and $10^{40}$ erg $s^{-1}$ and a dipole field component in the range of $10^{12}$ G $\leq$ $B_{dipole}$ $\leq$ $10^{13}$ G was suggested considering the observed luminosity. They reported the presence of a sinusoidal pulse shape with large variations in amplitude. According to their findings the system contains a massive OB giant or supergiant donor. In addition, \cite{2020ApJ...895..127B} found that ULX-7 has a super$-$orbital period as 38 days. Moreover, they identified a new transient ULX, XT-1, reaching a peak luminosity of $10^{40}$ erg $s^{-1}$.

The M51 galaxy hosts a large number of transient X-ray sources: One of them is the well-known source ULX-4. The first detection of ULX-4 was reported by \cite{2004ApJ...601..735T}. They presented that the flux value of ULX-4 (source 37, in their Table 2) increased by two orders of magnitude within the energy range (0.5$-$8 keV) from the analysis of 2000 and 2001 {\it Chandra} data. Also, ULX-4 is one of 8 ULXs catalogued by \cite{2011ApJ...741...49S} in this galaxy pair, and its unabsorbed luminosity is $L_{\mathrm{X}}$ = 2.2 $\times$ 10$^{39}$ erg s$^{-1}$ in the ($ 0.3-$ 10) keV energy range. Recently, \cite{2018MNRAS.476.4272E}, \cite{2020MNRAS.491.1260S} and \cite{2020ApJ...895..127B} reported the long-term flux variability of some of the ULXs in M51 including ULX-4 using {\it XMM-Newton}, {\it Chandra} and {\it Swift-XRT} observations.

In the present work, we focus on determining the nature of the transient source ULX-4 and its potential optical counterpart. Toward these goals, we carried out comprehensive spectral and temporal analyses of the X-ray observations of ULX-4 using the archival {\it XMM-Newton}, {\it Chandra} and {\it Swift} observations. 
A number of these observations have not been used in previous studies of this source. In addition, we make extensive use of archival {\it HST} observations in an effort to isolate and identify possible optical counterparts of ULX-4.

The paper is organized as follows:
Section \ref{sec:2} contains information on the X-ray and optical observations together with the details of the data reduction, along with the methodology deployed in the analysis of these data. In Section \ref{sec:3}, we present and discuss the main results. Finally, in Section \ref{sec:4}, we conclude by summarizing our major findings.

\section{OBSERVATIONS, DATA REDUCTION AND ANALYSIS} \label{sec:2}
\subsection{X-Ray Data}

\begin{figure*}
\begin{tabular}{ccc}
\includegraphics[scale=0.19]{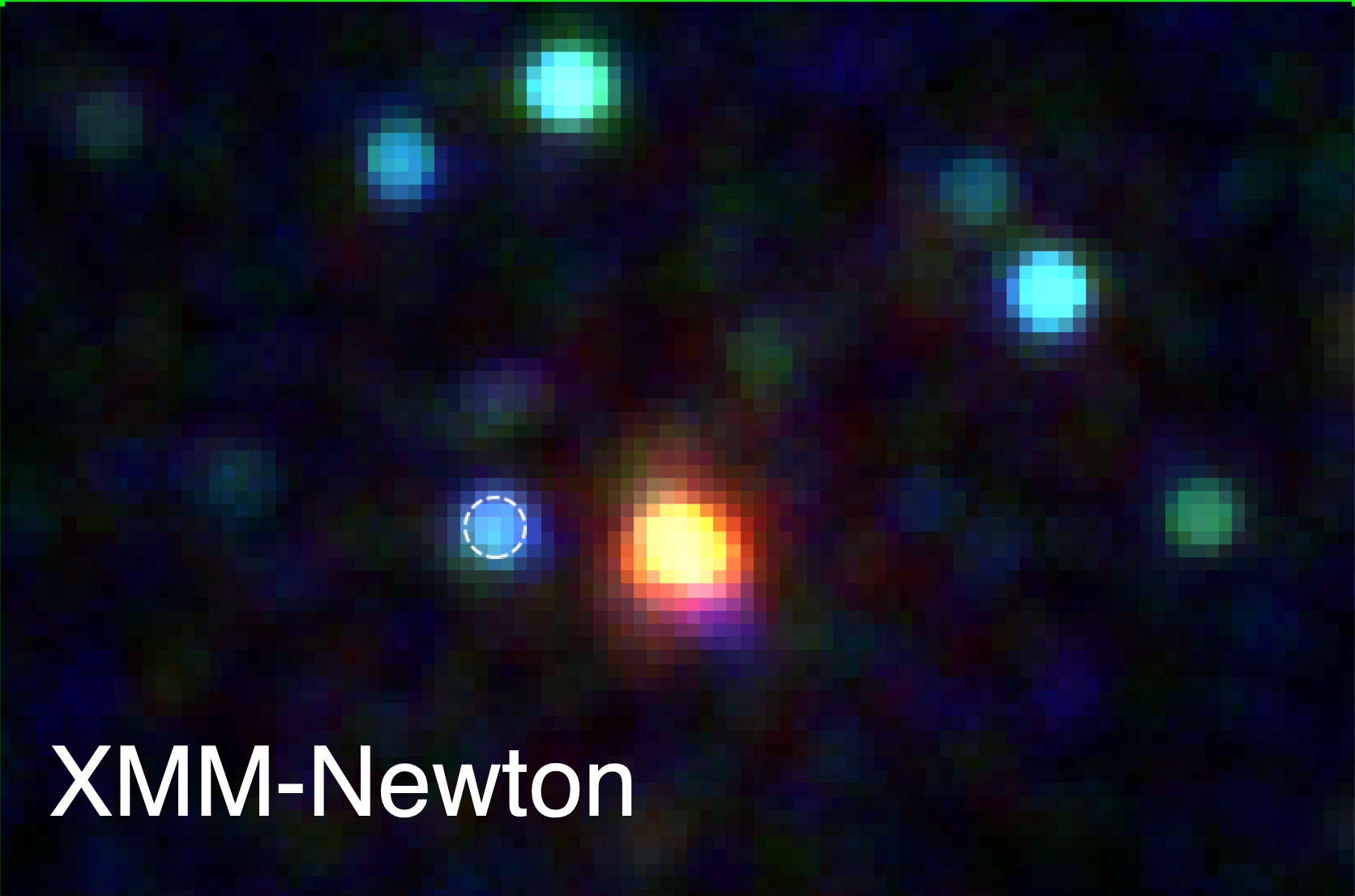}&
\includegraphics[scale=0.19]{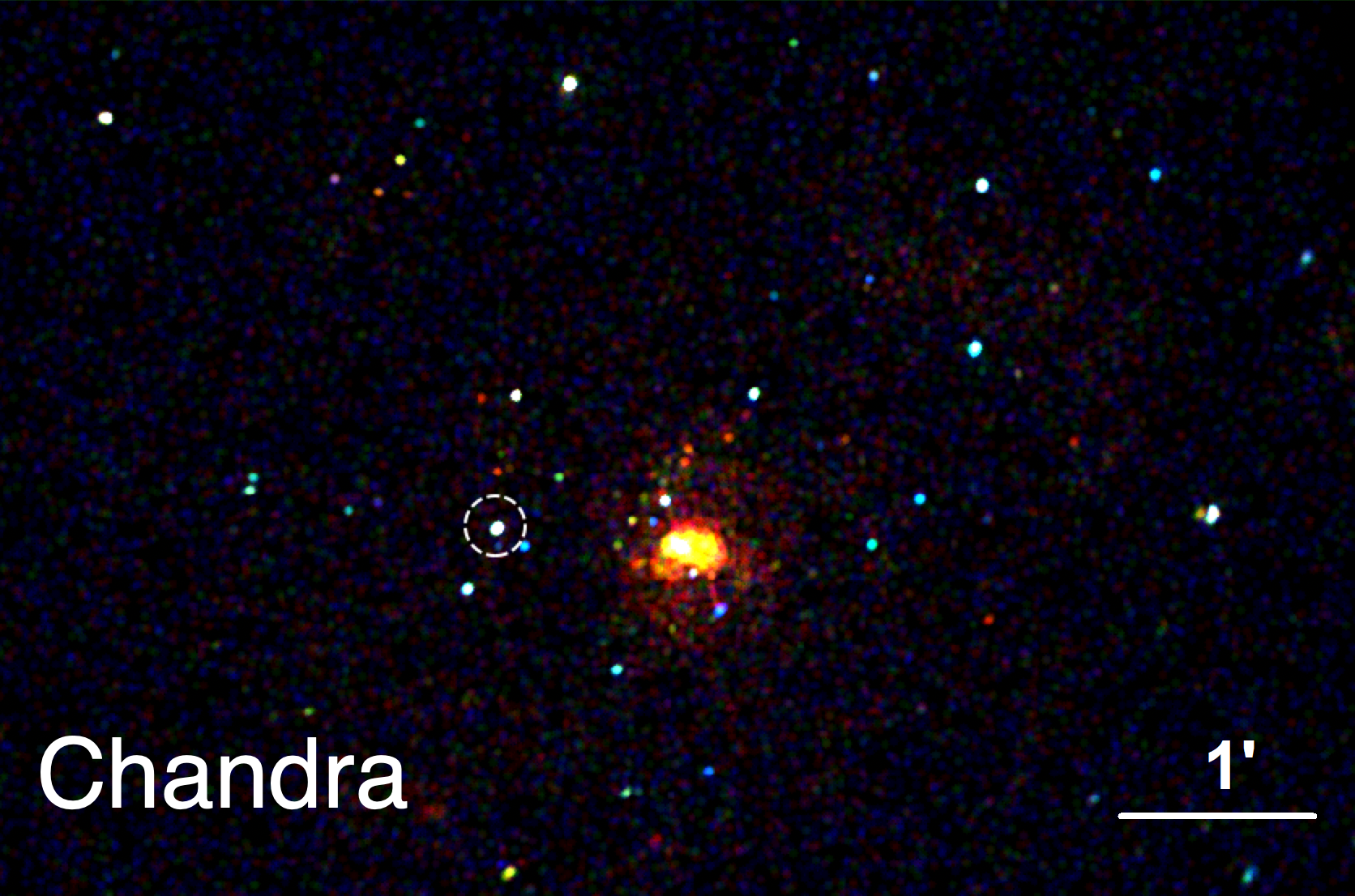}&
\includegraphics[scale=0.19]{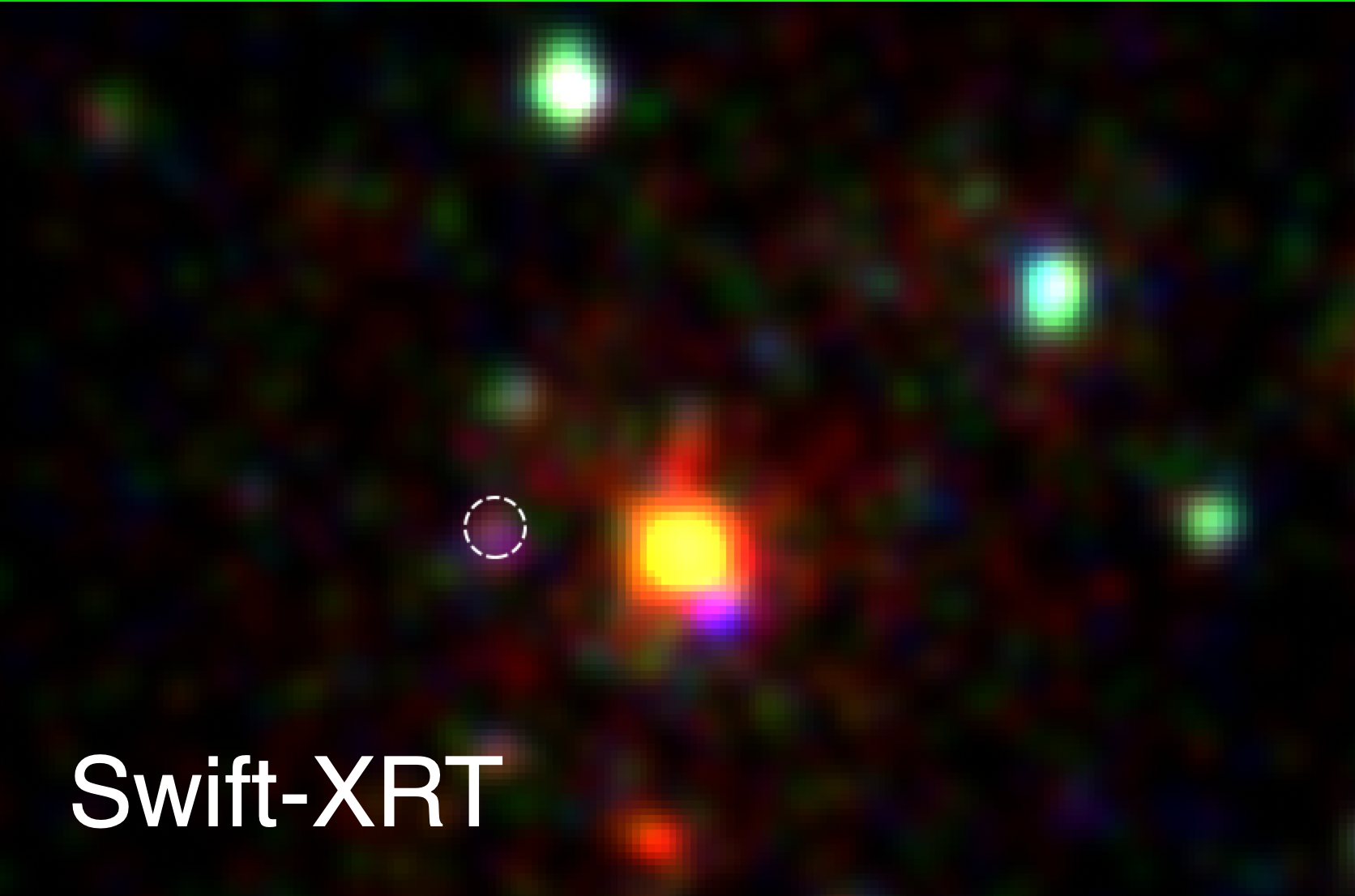}
\end{tabular} 
\caption{X-ray true color (Red :0.3-1 keV, Green: 1-2 keV and Blue: 2-7 keV) images of ULX-4. left panel: {\it XMM-Newton} (XM5), center panel:{\it Chandra} (C12) and right panel:{\it Swift-XRT} (stacked images: observations of Target ID 30083). The images were smoothed with a $3\arcsec$ Gaussian. The source ULX-4 is indicated by dashed white circles. All three images show the same region (4.9 $\arcmin$ $\times$ 7.4 $\arcmin$). North is left and east is up in all panels.}
\label{F:1} 
\end{figure*}

The source (ULX-4) was observed multiple times with {\it XMM-Newton}, {\it Chandra} and {\it Swift-XRT} over 20 years. Log of {\it XMM-Newton} and {\it Chandra} observations used are given in Table \ref{T:1}. The true color images of the source from the {\it XMM-Newton}, {\it Chandra} and {\it Swift-XRT} are shown in Fig.\ref{F:1}.

M51 ULX-4 was observed by {\it Swift-XRT} multiple times (276 observations) between 2005 and 2021. The target IDs of the observations we used in this study are 10717, 11417, 32017, 37267, 81964, 11106, 30083, 35895, 80113 and 745090. These observations have very different exposures. For example, the minimum exposure time is 5 s and maximum exposure time is 6857.3 s. Mostly the exposure times of the observations are around 1500 s.

The {\it XMM-Newton} EPIC data were analyzed with Science Analysis System ({\scshape sas})\footnote{https://www.cosmos.esa.int/web/xmm-newton/sas} v17.0 software.
The {\it epchain} and {\it emchain} tasks were used to obtain EPIC-pn and MOS event files for each observation. The events corresponding to PATTERN$\le$12 and PATTERN$\le$4 with FLAG=0 were selected for EPIC-MOS and pn cameras, respectively. We followed the standard procedure of extracting the source and the background events by deploying a circular radius of 15$\arcsec$ with the {\it evselect} task. We note though that if we compare {\it XMM-Newton} and {\it Chandra} images, it is clear that in most of the {\it Chandra} observations, that a number of faint point sources are located near ULX-4 and fall within the radius of 15$\arcsec$ region. This invariably contaminates the source counts for ULX-4.

{\it Chandra} ACIS-S and ACIS-I observations were analyzed by using Chandra Interactive Analysis of Observations ({\scshape ciao})\footnote{https://cxc.harvard.edu/ciao/} v4.12 software and calibration files {\scshape caldb} v4.9. The level 2 event files were obtained with {\it chandra\_repro} in {\scshape ciao}. The source and the background events were extracted from circular regions of 5$\arcsec$ radius. We obtained the source spectra and light curves with the tasks {\it specextract} and {\it dmextract}, respectively.

The count rates for Swift-XRT in PC mode were extracted using automated procedures \citep{2009MNRAS.397.1177E} specified on the web page\footnote{https://www.swift.ac.uk/user\_objects/}. In PC mode, source counts were extracted from a radius of 30 pixel circular region centered on the source position. The background counts were extracted from an annulus centered on the source with inner and outer radii of 60 pixels (142$\arcsec$) and 110 pixels (260$\arcsec$), respectively. As noted in the study of \citep{2009MNRAS.397.1177E}, some sources falling into the background region are excluded from the background estimate. We chose the default value of 20 counts/bin which the minimum number of counts a bin must contain to be considered complete in compliance with the guidelines indicated on the aforementioned website. When we applied these procedures, we used 82 observations with total counts greater than zero in the 0.3-10 keV energy band. 
ULX-4 has been observed four times with {\it NuSTAR} between 2017 and 2019. However, the source was not spatially resolved in these observations.

\subsubsection{Spectral Fitting}

The spectral properties of ULX-4 have been studied by several authors using {\it XMM-Newton} and {\it Chandra} data (e.g. \citealp{2005ApJ...635..198D, 2006ApJ...645..264T}). In this study, C13, C14, XM6, XM7, XM8 data sets were used for the first time for spectral analysis. In addition, the data sets used in previous studies were reanalyzed in detail. The ULX-4 spectra with the instrument responses and ancillary files were generated for {\it XMM-Newton}, {\it Chandra} and {\it Swift-XRT} observations. The spectral fitting was performed with {\scshape xspec} v12.8.2 for each observation. According to the total source counts, the source energy spectra were grouped with the FTOOLS {\it grppha} at least 10 and 30 counts per energy bin for {\it Chandra} and {\it XMM-Newton} data, respectively.

As discussed in several studies, one-component models should be considered statistically when the data quality is low and exposure is short. However, these models tend not to provide sufficiently accurate descriptions of the ULX spectra and hence the interpretation of the physical origin of the spectra remains questionable \citep{2009MNRAS.397.1836G,2013MNRAS.435.1758S,2017ARA&A..55..303K}. Although we do not have high quality data, we investigated single and two component models which are widely adopted in the literature \citep{2009MNRAS.397.1836G,2017A&A...608A..47K,2020AAS...23523702E,2021MNRAS.501.1002W}.
The ULX-4 spectra were fitted with single component models such as a {\it power-law},  diskblackbody ({\it diskbb}), pfree diskblackbody ({\it diskpbb}), broken power-law ({\it bknpower}) and Comptonization ({\it compTT}) model. We did attempt two component models such as {\it power-law+diskbb}, {\it power-law+mekal}, {\it power-law+diskpbb}, {\it diskbb+compTT} and {\it diskbb+cutoffpl}.  In addition to these models, we included two absorption components ({\it tbabs}).
One of the absorption models represented the line-of-sight column density, which we kept fixed at the Galactic value N$_{H}$ =3$\times$$10^{20}$ cm$^{-2}$ \citep{1990ARA&A..28..215D} and the other was left free to account for intrinsic absorption. We found that the intrinsic absorption values are small enough (between $10^{13}$ - $10^{17}$ cm$^{2}$) to be negligible for all observations. Therefore, a fixed column density (Galactic N$_{H}$) was used throughout this work.

In spectral analyses of the {\it XMM-Newton} and {\it Chandra} data, the spectral fits obtained show a wide range of $\chi^2_{\nu}$ values, therefore, the single and two component models were examined according to the $\chi^2_{\nu}$ values in the range of 0.8 <$\chi^2_{\nu}$<2 and F-test results. The {\it power-law}  model fits are statistically better than  the other single component models at a 3$\sigma$ confidence level for the {\it XMM-Newton} and {\it Chandra} datasets. 
On the other hand, using two-component models, no statistically significant improvement was achieved in fitting of the datasets, except for XM1 data. Only the spectrum of XM1 gives better statistics according to the F-test for the {\it power-law+mekal} model at 3$\sigma$, along with well constrained model parameters and $\chi^2_{\nu}$ statistics. Also, for the majority of the datasets, the parameters of the two-component models do not appear to be well constrained, although they have statistically acceptable fits in terms of reduced $\chi^2_{\nu}$ interval.

The values given in Table \ref{T:2} show the best model parameters. With the exception of the XM1 dataset, the parameters of the two-component models for the majority of the other datasets are unlikely to be meaningful considering the large uncertainties although the $\chi^2_{\nu}$ are in the range of 0.8 to 2, a range on the border of being acceptable.

The unabsorbed fluxes of ULX-4 were calculated in the energy band 0.3$-$10 keV using the convolution model {\scshape cflux} available in {\scshape xspec} and the source luminosity values were calculated assuming a distance of 9 Mpc. The {\it power-law} model does not have a natural cut-off at low energies, therefore the unabsorbed source fluxes can be overestimated. Considering this, we calculated the absorbed fluxes for observations with sufficient data statistics using the power law model. However, we found that there are negligible differences between the absorbed and unabsorbed flux values for all datasets. Therefore, only the unabsorbed fluxes are listed in Table \ref{T:2}.

In Table \ref{T:2}, the well$-$fitted spectral model parameters for ULX-4 obtained with {\it XMM Newton} and {\it Chandra} data are given with unabsorbed flux values. 
Three (XM6, XM7 and XM8) out of eight {\it XMM-Newton} observations we used do not have EPIC-MOS data.  
There appears to be too much noise in the MOS data for the majority of datasets, therefore, simultaneous fits of EPIC pn + MOS spectra do not produce acceptable model parameters and their chi-square statistics do not improve (except for XM1).

\subsubsection{Short-term X-ray Variability}

We searched for short-term variability of ULX-4 with recently available data of XM6, XM7, XM8, C13 and C14. The background subtracted X-ray light curves of ULX-4 were binned over intervals of 100s, 500s, 1000s and 2000s in the 0.3$-$10 keV energy band using {\it evselect} in {\scshape sas}. The resulting light curves were tested for short-term count variations using $\chi^2_{\nu}$ test. This test was applied to search for large amplitude variations
with respect to the constant count rate hypothesis. The results of $\chi^{2}$ test were examined together with the probability values P(var), an indicator of the variability of data. If P(var) exceeds 95\% taken as the limit, it could be interpreted as data variability. However, we found a P(var) value of $\sim$ 57\%. This moderate value suggests that ULX-4 does not show any significant amplitude variation.

In addition, the background subtracted X-ray light curves, sampled at 0.1 s, yielding 5 Hz Nyquist frequency, were Fourier transformed to generate power density spectra (PDS) using the utility
{\scshape xronos} v6.0 in {\scshape heasoft} v6.27. Initially, the entire duration of the light curves was included in the transform to look for overall variability features but none were found. Then, the total exposure was divided into eight time intervals, and the (transform) procedure was repeated. No signal stronger than 1.5$\sigma$ significance was detected.

\subsubsection{Long-term X-ray Variability}

We also probed the long-term variability of ULX-4 by deploying the more recent data adding to previous studies of \cite{ 2018MNRAS.476.4272E}, \cite{2020MNRAS.491.1260S}, and \cite{2020ApJ...895..127B}. The background subtracted source count rates for {\it XMM-Newton} and {\it Chandra} data were obtained for the light curve in 0.3$-$10 keV energy band using {\scshape xspec}. The background subtracted source count rates of {\it Swift-XRT} in PC mode were extracted using standard procedures. We only used 82 data sets, between 2005-2021 years, from which the source was detected or 3$\sigma$ upper limits could be obtained.
All these count rates were converted to fluxes with the the best fitting parameters of {\it power-law} model $\Gamma$ = 1.75 and N$_{H}$= 3$\times$ 10$^{20}$ cm$^{-2}$ using the {\it WebPIMMS} tool\footnote{https://heasarc.gsfc.nasa.gov/cgi-bin/Tools/w3pimms/w3pimms.pl}. In addition,
the count rates for C1 and C3 data sets were obtained by using {\it srcflux} tool in {\scshape ciao} with the same $\Gamma$ and N$_{H}$ for 3$\sigma$ upper limit. The resultant long-term light curve of ULX-4 is given in Fig.\ref{F:3}.
The {\it XMM-Newton} and {\it Swift-XRT} detectors do not have sufficient spatial resolution to resolve sources too close to ULX-4. Although we chose a small source extraction region (radius of 15$\arcsec$) for {\it XMM-Newton} data, there are still three faint transient X-ray sources detected within this region. To calculate the contributions of these three sources, we derived their fluxes using all {\it Chandra} observations given in Table \ref{T:1}. For this, CIAO's {\it srcflux} task was used by taking the photon index of $\Gamma$=1.7 and N$_{H}$=0.03$\times$10$^{22}$. We calculated the maximum total flux from the contaminants as 2.29$\times$10$^{-14}$ erg cm$^{-2}$ s$^{-1}$. This potentially implies that the {\it XMM-Newton} and {\it Swift-XRT} fluxes of ULX-4 are overestimated by this amount due to the presence of these nearby sources.  However, we need to keep in mind that these sources are transient and the observations are not simultaneous.

We constructed Lomb-Scargle (LS) periodograms \citep{1976Ap&SS..39..447L,1982ApJ...263..835S} to search for periodic modulations of the ULX-4 system for the long-term light curve by using a {\it Python}\footnote{https://docs.astropy.org/en/stable/timeseries/lombscargle.html} subroutine.
However, no periodicity was detected in the period range 2-3000 days from any of the X-ray data. In addition, we re-ran the LS analysis on subsets of close observations. For this, we divided the light curve into six epochs. These epochs were defined as Epoch 1: 2005, Epoch 2: 2007, Epoch 3: 2011, Epoch 4: 2012, Epoch 5: 2018-2021, Epoch 6: 2011+2012. However, we did not find any significant periodicity for these epochs.

\subsubsection{Time Lags}
Evidence for the presence of spectral lags (both hard and soft) in a number of variable ULXs including NGC 5408 X-1 \citep{2010MNRAS.405L..86H,2013MNRAS.436.3782D,2015MNRAS.453.2877H}, NGC 55 ULX-1 \citep{2017MNRAS.468.2865P}, NGC 1313 X-1 \citep{2020MNRAS.491.5172K}, and NGC 4559 X-1 \citep{2021MNRAS.504..551P} has been reported. These lags are thought to involve the accretion flow and strong outflow winds in which the hard photons propagate through an extended optically thick medium, the origin of which is speculated to be the base of the outflows. Another possible scenario, proposed by \cite{2013MNRAS.433.3453D} for hard lags, is that it represents the light-travel time between the corona and/or the neutron star surface or boundary layer, and the innermost region of the accretion disc. In X-ray binaries hard lags are usually seen for the upper kHz QPOs for which the frequency of either the lower or the upper kHz QPO is equal to the Keplerian frequency at the inner edge of the accretion disc (e.g. \cite{1998ApJ...508..791M,1999ApJ...522L.113O,1999PhRvL..82...17S}). As ULX-4 indicates a relatively high degree of variability based on the hardness-intensity diagram, we have attempted an extraction of frequency-dependent time lags following the procedure outlined in \cite{2014A&ARv..22...72U} and \cite{2021MNRAS.504..551P}. We used background subtracted {\it Chandra}- (C2, C4, C11, C12, C14) light curves in the 0.3–2 keV and 2–10 keV energy bands for which the source counting statistics are comparable. We also extracted lags for the following bands: 0.3 – 1.0 keV and 1.0 – 7.0 keV. We performed FFT for the full range of the two light curves binned with $\Delta$T=100 s over 3 to 15 intervals depending on the duration of the light curve (with each segment being 5 ks in duration). The resulting power spectra were averaged and then cross-correlated to obtain the spectral lag as a function of frequency \citep{2019JOSS....4.1393H,2019ApJ...881...39H}. We summarize our results as follows: the weighted spectral lags are consistent with zero within error bars for both sets of the energy bands. We note, in passing, that if we follow the procedure of \cite{2021MNRAS.504..551P} i.e., taking the average lag over a certain narrow frequency range, for which the coherence is relatively high, we too find a hard lag for one of the light curves (C2); the value is moderately high (758 +/- 317 s) but is quite sensitive to the chosen frequency range(2.0E-4 – 4.4E-4) and the average coherence level (0.33).\\

In order to assess the statistical significance of the extracted time lag, we performed a Monte Carlo simulation of 1000 light curves and extracted frequency-resolved lags and the associated coherence. The simulations were setup with the mean count-rate and rms variability parameters estimated from the {\it Chandra} light curves for the 0.3–2.0 keV and 2.0 – 10 keV bands. We used the broken-power law (BPL) as the spectral model with nominal parameters. The resulting light curves (25 ks in duration and binned at 100 seconds) were analyzed and frequency-resolved time lags were extracted using the {\it STINGRAY} package following the procedure outlined in \cite{2014A&ARv..22...72U} and \cite{2021MNRAS.504..551P}. We found the mean time lags and the coherence to be consistent with zero within the uncertainties. As another check, we also averaged the mean lags over a narrow frequency range (i.e., (2–4.4)$\times10^{-4}$ Hz) for a number of individual light curve combinations for which the coherence was found to be in range $\geq$ 0.4 and above. We obtained the average to be 349 $\pm$ 189 s for one of the light curve combinations. In totality, approximately 60 light curve-combinations from our of sample of 1000 light curves (over two energy channels each) yield a similar result thus suggesting that a marginally significant lag can be obtained spuriously (with a relatively high probability of $\sim$ 6\%) over a limited frequency range. We thus urge caution in attaching a physical significance to the experimentally extracted lag noted above.   

\subsubsection{New Transient X-ray Source: CXOU J132951.7+471010}
In the process of analyzing the {\it Chandra} observations from 2017, we identified a new transient source. As shown in Fig.\ref{F:5}, this transient source, shown in panel C13 as a white dash circle, is not seen in the other available {\it XMM-Newton} and {\it Chandra} observations. We determined that the source is located at the coordinates R.A (J2000) = 13$^{\mathrm{h}}$29$^{\mathrm{m}}$51$^{\mathrm{s}}.7$, Dec. (J2000) = +47$^{\circ}$10$\arcmin$10.$\arcsec10$ (see Fig.\ref{F:5}) using {\scshape ciao} {\it wavdetect} tool and it has not been cataloged by \citet{2016ApJ...827...46K}. In addition, we were unable to find this source in the Transient Name Server\footnote{https://wis-tns.weizmann.ac.il} among the transients in M51. For convenience we use the acronym CXOU (for {\it Chandra} X-ray Observatory Unregistered source) and name this source as CXOU J132951.7+471010. Its count rate is found as (9 $\pm$ 3.12) $\times 10^{-4}$ count/s in the 0.3$-$10 keV energy range from 2017 {\it Chandra} data. Flux upper-limits for non-detections were also derived at 3$\sigma$ assuming a {\it power-law} spectrum with $\Gamma$ = 1.7 and N$_{H}$ = 3$\times$ $10^{20}$ cm$^{-2}$ using {\scshape srcflux} tool in {\scshape ciao}. {\it Chandra} flux was obtained with help of {\it WebPIMMS} tool by using the same parameters. We obtained the unabsorbed flux of CXOU J132951.7+471010 in the range of (0.05$-$1.88)$\times$10$^{-14}$ erg cm$^{-2}$ s$^{-1}$ and the corresponding luminosity of (0.05$-$2)$\times$10$^{38}$ erg s$^{-1}$ assuming a distance of 9 Mpc.

\subsection{Optical Data}

\subsubsection{Astrometry}
\label{Astrometry}
We used {\it HST} archival data (see Table \ref{T:3} for details) to investigate the optical properties of ULX-4. For a relative astrometry between {\it Chandra} (ObsID 13816) and {\it HST} (ObsID J97C51R4Q) images, we followed a similar method that we used in our previous works \citep{2020MNRAS.499.5682A, 2021MNRAS.505..771E}.
The {\scshape daofind} tool in {\scshape iraf}\footnote{http://ast.noao.edu/data/software} and {\it wavdetect} tool in {\scshape ciao} were used for source detection in {\it HST} and {\it Chandra} images, respectively. In both images, three isolated and point reference sources were matched; their coordinates and counts are given in Table \ref{T:4}. These reference sources are (1) the SN 2011dh \citep{2013MNRAS.436.1258H}, (2) an X-ray binary (CXOU J133006.5+470834) \citep{2016ApJ...827...46K} and (3) a background radio source (J133011+471041) \citep{2015MNRAS.452...32R}. We derived the offsets between these {\it Chandra} and {\it HST} reference sources as 0\farcs21, 0\farcs17 and 0\farcs18, respectively at 90\% confidence level. The astrometric errors between the {\it Chandra} and {\it HST} images were found 0\farcs14 for R.A. and 0\farcs11 for Dec. The astrometric correction was calculated using quadratic sum for the standard deviations of these errors. As a result, we found the position of the ULX-4 on the {\it HST} image within an error radius of 0\farcs18 at 95\% confidence level. We identified two optical counterparts within the error radius and labelled them as A and B according to their increasing Dec. The corrected coordinates of ULX-4 are also given in Table \ref{T:4}. The {\it HST} three color (red, green and blue; RGB) image showing the position of ULX-4 and its optical counterparts (A and B) is depicted in Fig.\ref{F:6}.

\subsubsection{Photometry}
\label{Photometry}
Point Spread Function (PSF) photometry was performed to determine the magnitudes of optical sources using the {\scshape doaphot} package \citep{1987PASP...99..191S} in {\scshape iraf}. In order to obtain photometric errors, we multiplied the pixel values by the exposure time using {\scshape imarith} tool to calculate the electron/pixel. Thirty bright and isolated sources near the optical counterparts for ULX-4 were selected to build the PSF model. PSF fitting radius was taken as 3 pixels in the {\it allstar} task. For ACS/WFC and WFC3 filters, the zero-points were obtained from {\scshape pysynphot}\footnote{https://pysynphot.readthedocs.io/en/latest/} which are given in Table \ref{T:5}. We performed aperture corrections with a radius between 0\farcs05 to 0\farcs5 for ACS/WFC and ACS/WFC3.

\subsubsection{Spectral Observations}
\label{spectra}
The obtained magnitudes of the optical counterparts were corrected for extinction using the ratios of Balmer lines of the region of ULX-4. For this purpose, spectral data were obtained from the TUG Faint-Object Spectrograph and Camera (TFOSC) instrument mounted on RTT150 (Russian–Turkish Telescope in Antalya, Turkey)\footnote{http://tug.tubitak.gov.tr/en/teleskoplar/rtt150-telescope-0} on 16 July 2018. The derived H$_{\alpha}$ and H$_{\beta}$ flux ratio for the spectrum have been taken into account to determine Balmer decrements. The standard data reduction steps were performed using {\scshape iraf}. For the reddening calculation, intrinsic Balmer decrements ratio (H$_{\alpha}$/H$_{\beta}$)$_{int}$ value was taken as the standard 2.87 for star forming galaxies, estimated using the temperature T = 10$^{4}$ K and electron density n$_{e}$ = 10$^{2}$ cm$^{3}$ for Case B of \cite{1989agna.book.....O}. E(B$-$V) was found as 0.15 mag. from the ratio of hydrogen Balmer lines. Then, A$_{V}$ was calculated as 0.46 mag using the equation of A$_{V}$ = R$_{V}$ $\times$ E(B$-$V), where the extinction factor of R$_{V}$ is 3.1 \citep{1989AJ.....97.1099C}. The extinction corrected magnitudes of the optical counterparts were calculated in accordance with A$_{V}$ given in Table \ref{T:5}. 

Also we found a single optical counterpart of the CXOU J132951.7+471010 source within the error radius of 0.$\arcsec$27 at 99.7\% confidence level. The optical counterpart of CXOU J132951.7+471010 is detected only in UV bands ($m_{F275W}=22.03\pm0.04$ mag and $m_{F336W}=22.88\pm0.03$ mag).

\subsubsection{Spectral Energy Distribution}

Assuming optical emission originates from the donor star, the spectral types of optical counterparts were estimated from Spectral Energy Distributions (SEDs) obtained by using {\scshape pysynphot}. The CK04 standard stellar spectra templates were used in this program \citep{2004A&A...419..725C}. The flux values of the counterparts of A and B were obtained from the magnitudes given in Table \ref{T:5}. All synthetic spectra were derived with metallicity of Z = 0.015 \citep{2018MNRAS.475.3561U} and extinction of A$_{V}$ = 0.46 mag. Also, synthetic spectra were normalized with Vega m$_{V}$ = 0 mag. In SED plots, the wavelength of the filters were selected as the pivot wavelength, obtained from {\scshape pysynphot}. By way of a comparison of the spectral features of the optical counterparts, with those of two field stars (R1 and R2 in Fig.\ref{F:6}), the SEDs of optical counterparts and selected field stars are shown in Fig.\ref{F:9}. 

In addition, as an alternative way to determine the spectral type of counterparts, we fitted a blackbody model to the SEDs of these counterparts. A code has been used with {\scshape optimset} and {\scshape fminsearch} functions in {\scshape matlab}\footnote{https://www.mathworks.com/matlabcentral/fileexchange/20129-fit-blackbody-equation-to-spectrum} to get a blackbody spectrum. The SEDs for the optical counterparts A and B are adequately fitted to a black-body spectrum with temperatures of 26554 $\pm$ 104 K and 21386 $\pm$ 173 K, respectively, at 95\% confidence level. The $\chi^2_{\nu}$ values were found 0.88 and 0.91, respectively. The number of degrees of freedom is four. Moreover, R1 and R2 are also well-fitted to a black-body spectrum with temperatures of 27200 $\pm$ 187 K and 17378 $\pm$ 214 K, respectively, at 95\% confidence level.

\subsubsection{Long-term Optical Variability}

In order to search for periodic modulations, we constructed long-term light curves. We used thirty {\it HST} ACS/WFC F606W and F814W observations (Proposal ID: 14704) with the same exposure time of 2200 s. The same procedure (as given in Section \ref{Photometry}) was followed for PSF photometry for all observations.

A moderate periodic modulation was determined for counterpart A in the F606W filter. No modulation was observed in the F814W filter. For counterpart B, no modulation with significant amplitude was found in either filter. To check whether the modulation in counterpart A could be due to systematic error, we obtained the light curves of 10 reference sources. These reference sources were selected as bright sources close to counterpart A within the white circle in Fig.\ref{F:6}. We did not find any periodic modulation in the reference sources. 
Then we calculated the average magnitude of the source using the magnitudes of 30 observations. We calculated the deviations of magnitude by taking the difference of the source's average magnitude from its magnitude value.
Afterwards, these deviations were subtracted from the magnitude of counterpart A for each observation which are given in Table \ref{T:6}. As a result, a periodic modulation was seen clearly from the corrected light curve of the counterpart A in the F606W images (Fig.\ref{F:10}). We fitted a sinusoidal curve using the equation m(t) = $\overline{m}$ + Asin[2$\pi$({\it t}-{\it t$_{1}$})/{\it P} + $\phi$] given by \cite{2009ApJ...690L..39L}. Here, {\it P} is a period, $\phi$ is a phase angle, {\it t} is the reference time, {\it A} is amplitude and $\overline{m}$ is an average magnitude. When we applied this sinusoidal fit to the light curve, {\it P} = 264.45$\pm$36.81 days, {\it A} = 0.07 $\pm$ 0.02 magnitude and $\phi$ = 96$^{\circ}$.35$\pm$0.78.
These values were obtained with $\chi^{2}$/dof = 21.06/27 at 99.7\% confidence level.

\subsubsection{Age and Mass Estimation of a Nearby Cluster}
It is known that the bright X-ray sources are located near star clusters and their physical parameters such as mass and age provide information about their nature \citep{2002ApJ...577.726}. In our study, optical counterparts of ULX-4 are also located near a star cluster with coordinates of R.A.= 13$^{\mathrm{h}}$29$^{\mathrm{m}}$53$^{\mathrm{s}}.278$ and Dec.= +47$^{\circ}$10$\arcmin$42$\arcsec.55$ cataloged by \cite{2016ApJ...824...71C}. In Fig.\ref{F:6}, the location of the cluster and optical counterparts are shown.
We derived the age and mass values of this cluster using the data obtained from ACS/WFC and ACS/WFC3 (see Table \ref{T:3}). For this calculation, we performed aperture photometry with the radius of 0.$\arcsec$4 using {\it apphot} package in {\scshape iraf}. We obtained magnitudes for F336W, F435W, F555W and F814W filters as 17.92$\pm$0.01 mag, 19.04$\pm$0.01 mag, 18.96$\pm$0.01 mag and 18.63$\pm$0.01 mag, respectively.

Simple Stellar Populations (SSPs) define the group/cluster of stars with the same age and chemical composition. SSPs are the most fundamental prediction of
population synthesis models to determine the spectral evolution of
stellar populations \citep{2003MNRAS.344..1000B}. Comparison of the integrated colors of individual clusters with SSPs gives an estimate of their ages. Here, we used $U-B$, $B-V$ and $V-I$ colors of the cluster and compared with the SSP models with the color excess E(B-V)=0.15 mag and metallicity of Z = 0.015.
$U-B$ and $V-I$ colors were dereddened using the extinction coefficients from \cite{2005PASP..117.1049S}. We derived the age of cluster as 6.73 Myr from SSP models.

The mass of this cluster was obtained from the extinction-corrected V-band luminosity and the age dependent mass-to-light ratio predicted by the SSP models with the metallicity and the distance modulus of 29.62 mag of the galaxy M51.
We selected two-part {\it power-law} IMF (Initial Mass Function) from the models and found a total mass of $\sim$ 3.2$\times10^{4}$ M$_{\odot}$ with the age of 6.73 Myr for this cluster.

In addition, we constructed color-magnitude diagrams (CMDs) to determine ages and mass of the optical counterparts. Padova Stellar isochrones (PARSEC; \citealp{2012MNRAS.427..127B}) were used in CMD. We estimated the age of counterparts A and B as 9 Myr and 10 Myr, respectively: The masses are constrained as 20 M$\sun$ for A and 18 M$\sun$ for B.

\section{Results and Discussion}
\label{sec:3}

We investigated the X-ray and optical properties of ULX-4 in the Whirlpool galaxy M51 using archival {\it XMM-Newton}, {\it Chandra}, {\it Swift-XRT} and {\it HST} data spanning over 20 years. In the following sections, we discuss our main results.

\subsection{X-Ray}
We reanalyzed the majority of archival {\it XMM-Newton}, {\it Chandra} and {\it Swift-XRT} data sets noted earlier. Some of these data have not been used for spectral modelling and timing analysis in previous studies. Among the applied spectral models, an absorbed power-law model provided statistically acceptable fits to the {\it XMM-Newton}, {\it Chandra} datasets with an index of 1.44$-$2.0.
This range  is consistent with the source being in a hard state characterized by non-thermal emission \citep{2006AAS...209.0705R,2009ApJ...696.1712F,2010ApJ...716..181J,2011ApJ...741...49S,2019A&A...621A.118K}. Assuming accretion at Eddington luminosities, the mass of the compact object would be in the range 5$-$15 M$\odot$ i.e., a stellar-mass BH.

The unfolded energy spectra of C12 data using {\it power-law} model is given in Fig.\ref{F:specraChandra}. The profile of the spectrum is similar to that described as a hard ultra-luminous state by \cite{2013MNRAS.435.1758S,2021AstBu..76....6F}. The luminosity versus photon index for ULX-4 is shown in Fig.\ref{F:GammaLX}.
To investigate the possibility of a correlation between $L_{X}$ and $\Gamma$, we determined the Spearman rank correlation coefficient ($\rho$). We find a $\rho$ value of -0.22 with a significance of 0.5. The magnitude of the coefficient is too small to suggest any significant correlation.

Nonetheless the source exhibits evidence for hard spectra especially for the {\it Chandra} data sets (black points). Although the uncertainties in the extracted $\Gamma$ are relatively large, the {\it XMM-Newton} data sets, on the other hand, seem to show somewhat softer spectra (blue points). We now examine this particular feature in the data: a number of {\it XMM-Newton} observations, notably, XM1, XM4 and XM5 (see Table \ref{T:2}), indicate the presence of the {\it mekal} component in the spectral fits in addition to the {\it power-law}. We used the F-test tool (in {\scshape xspec}) to determine whether the addition of this new component led to any improvement of the fits. We found F-test probabilities as $10^{-5}$, 0.17 and 0.23 for XM1, XM4 and XM5, respectively. These values indicate that the {\it power-law+mekal} model provides a statistically significant improvement only for XM1 data. Unfolded spectra of XM1 is shown in Fig.\ref{F:XM1_uns}. We determined that the fraction of mekal component in the total flux from {\it power-law+mekal} is $\sim$10\% in the 0.3–2 keV band for these three observations.

The presence of the {\it mekal} component could be in part due to emission originating from a hot diffuse thermal plasma \citep{2005ApJ...633.1052F, 2017A&A...608A..47K}. It is also notable that {\it Chandra} observations indicate the presence of point-like sources near ULX-4 which are not necessarily spatially resolved in the {\it XMM-Newton} data (see Fig.\ref{F:5}). These sources could add to the background seen in the ULX-4 spectra. In order to further probe whether the soft excess emission is intrinsic to ULX-4, we compared the excess flux in the 0.3–2 keV band modeled as a {\it power-law}+{\it mekal}, with the diffuse emission from its surrounding region. We estimated the flux of the diffuse emission from regions nearby ULX-4. For this, the {\it mekal} temperature of kT as 0.6 keV (as given in Table \ref{T:2}) was taken as an input for {\it WebPIMMS} tool. We found that the soft excess emission is comparable to the nearby diffuse emission. This implies that the soft excess emission is unlikely to be associated with ULX-4. This interpretation is consistent with the study of \cite{2005ApJ...635..198D} using the 2000-2001 {\it Chandra} and the 2003 {\it XMM-Newton} data for other ULXs (except source 69; ULX-7) in M51.

\begin{figure}
\begin{center}
\includegraphics[width=\columnwidth]{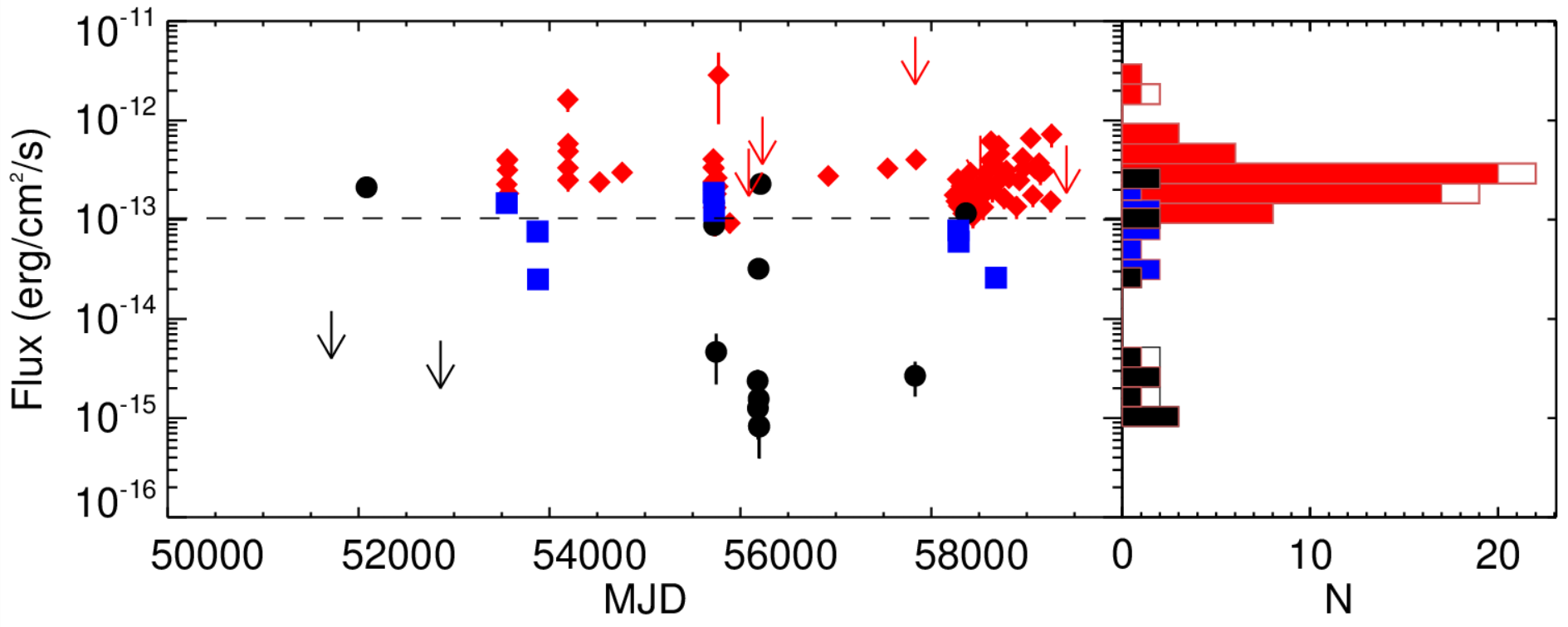}
\caption{The long-term X-ray light curves of ULX-4 (left) and the histogram panel (right). {\it Chandra}, {\it XMM–Newton} and {\it Swift-XRT} observations are marked filled black circles, blue squares and red diamonds, respectively. 3$\sigma$ upper limits for {\it Chandra} and {\it Swift-XRT} data are shown by the black and red downward arrows, respectively and also they are shown with open squares in the histogram}. For the histogram {\it Swift-XRT} observations are binned with 10-day binning. N is the number of observations. Dashed black line indicates the flux threshold for ULX state.
\label{F:3}
\end{center}
\end{figure}

\begin{figure}
\begin{center}
\includegraphics[width=\columnwidth]{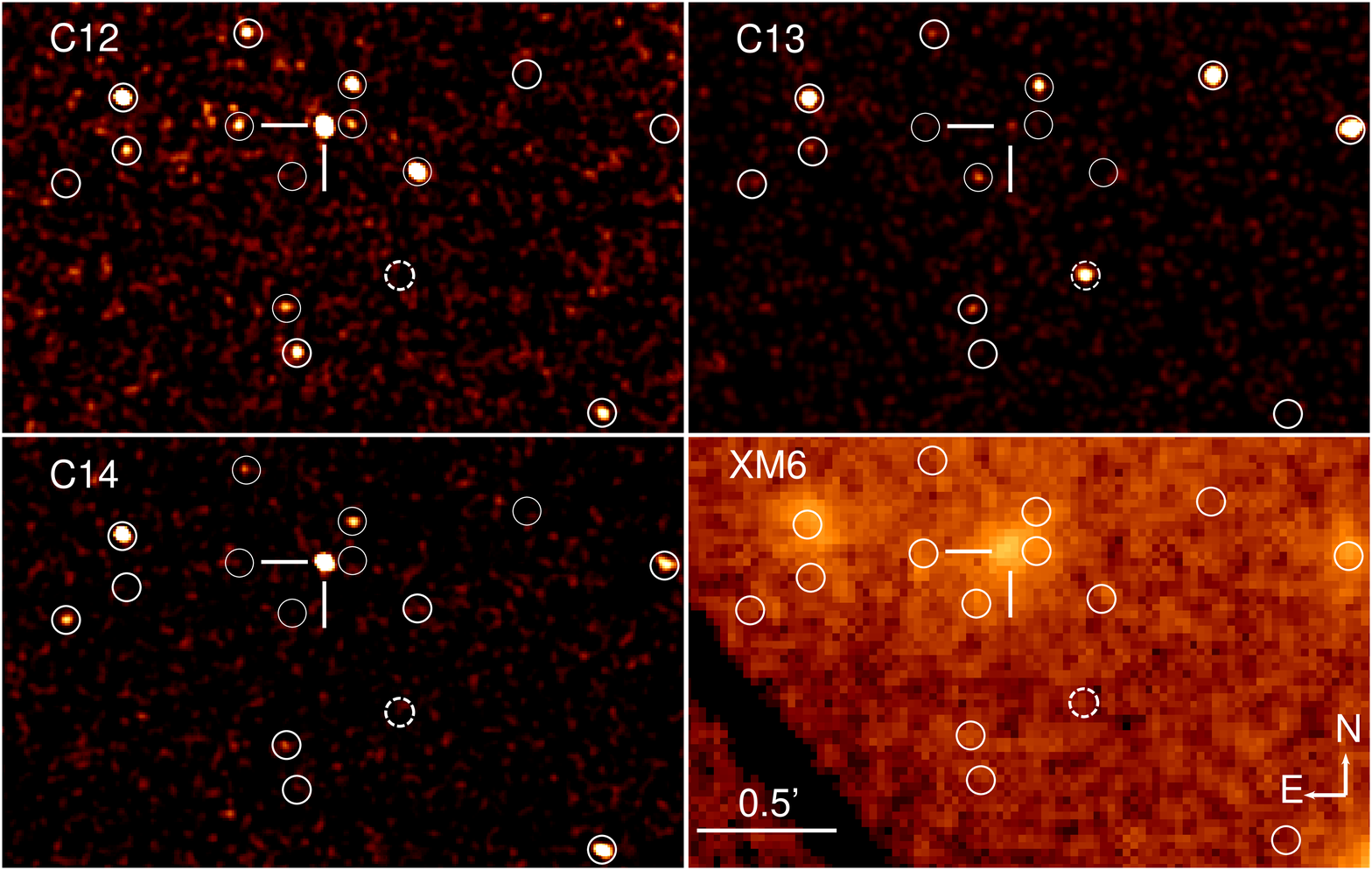}
\caption{Comparisons of ULX-4 (white bars) and new transient source CXOU J132951.7+471010 (dashed white circle) in C12 (2012), C13 (2017), C14 (2018) and XM6 (2018) observations, respectively. White circles with a radius of $3\arcsec$ represent some of the X-ray sources cataloged by \citet{2016ApJ...827...46K}. All panels were obtained with the same scale and the images were Gaussian smoothed with a radius of $3\arcsec$.}
\label{F:5}
\end{center}
\end{figure}

\begin{figure}
\begin{center}
\includegraphics[width=\columnwidth]{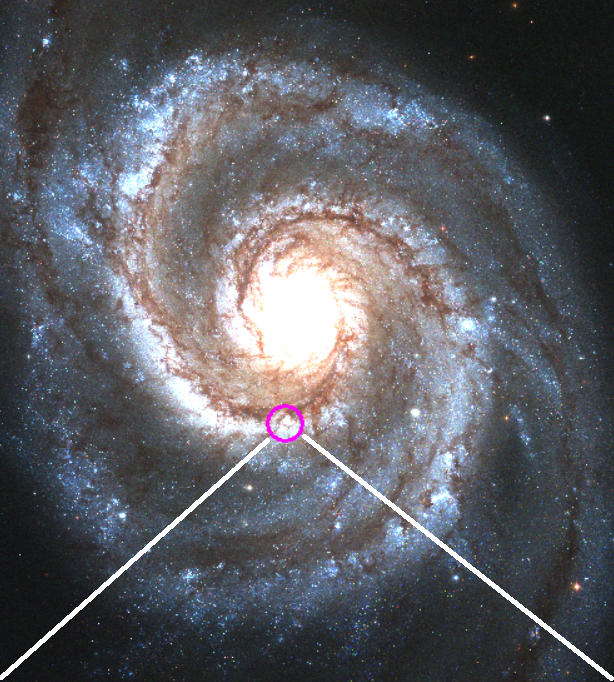}
\includegraphics[width=\columnwidth]{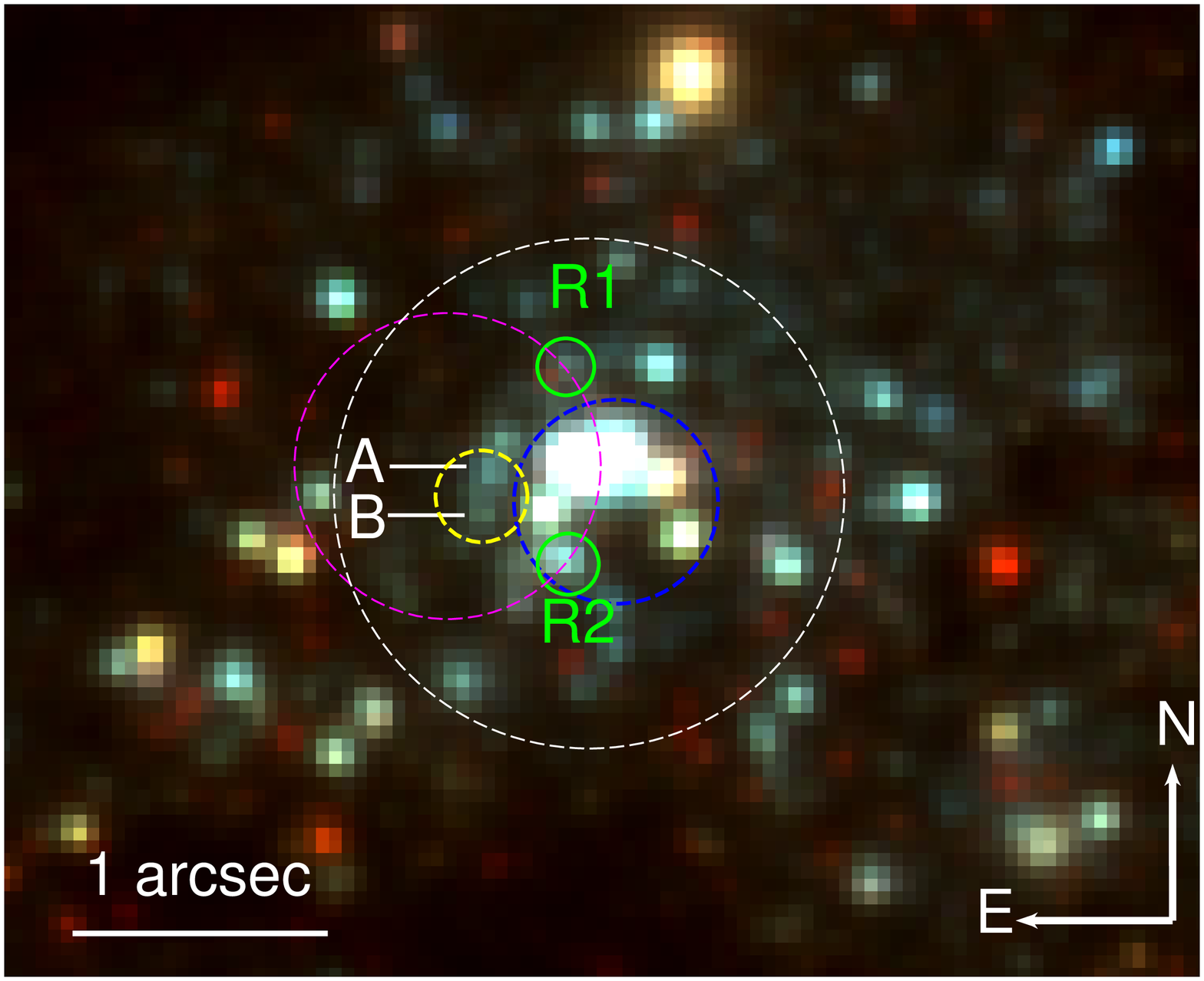}

\caption{The {\it HST} RGB image (R:F814W, G:F555W and B:F435W) of ULX-4 and its environment. The dashed magenta circle represents the {\it Chandra} position of ULX-4 with a radius of 0$\farcs$6. The dashed yellow circle represents the astrometric corrected coordinates of ULX-4. The optical counterparts A and B are also shown with white bars. The dashed blue circle indicates the star cluster. The dashed white circle represents the region used for reference sources with a 1$\arcsec$ radius. Also, two field stars (R1 and R2) are shown with green circles which are used for SEDs.}
\label{F:6}
\end{center}
\end{figure}

\begin{figure}
\begin{center}
\includegraphics[width=\columnwidth]{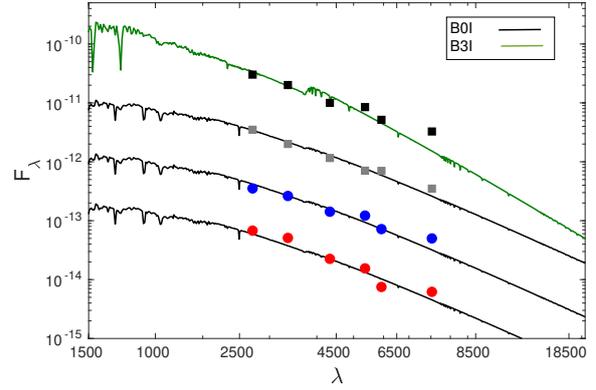}
\caption{The SEDs of optical counterparts A and B with field stars R1 and R2. The black and green lines represent the synthetic spectra for B0I and B3I, respectively. The red and blue filled circles represent flux values of counterparts for A and B, respectively. The filled black and gray squares represent flux values of sources for R1 and R2, respectively. All SEDs were derived with metallicity of Z$_{\odot}$ = 0.02 and extinction of A$_{V}$ = 0.46 mag. Systematic errors are less than 4\%. The lines were shifted upward by factor of 10 for clarity. The units of y and x axes are erg s$^{-1}$ cm$^{-2}$ and \AA, respectively.}
\label{F:9}
\end{center}
\end{figure}

Using majority of the archival observations of {\it XMM-Newton}, {\it Chandra} and {\it Swift-XRT}, we obtained the long-term X-ray light curve of ULX-4 (see the left panel of Fig.\ref{F:3}).

As seen from Table \ref{T:2}, the spectral fitting results of data, we calculated the
X-ray variability factor ($V_{f}$ = $F_{max}$/$F_{min}$) of ULX-4 as $\sim$ 230. 
Here, $F_{max}$ and $F_{max}$ represent maximum and minimum of flux values, respectively. This factor was obtained by using the 2012 {\it Chandra} observations (C6 $-$ C12) taken within a month. The source does not show significant variability from previously unused {\it Swift-XRT} observations (between MJD 57003 (2015) and MJD 59195 (2021)). We note that the propeller effect has been mentioned as a possible mechanism for the occurrence of high flux variability in ULXs (e.g. \citealp{2016MNRAS.457.1101T,2018MNRAS.476.4272E,2020ApJ...891..153E}).

\cite{2018MNRAS.476.4272E} and \cite{2020MNRAS.491.1260S} have noted the existence of a bi-modal flux distribution in some ULXs, including also ULX-4. We have tested this scenario for ULX-4 by using the available data, and we find some evidence, in the {\it Chandra} data, for such a distribution (see the right panel of Fig.\ref{F:3}). We note at the same time that the statistics are very low and that the bi-modal feature is not evident in either the {\it XMM-Newton} or the {\it Swift-XRT} data sets. If the bi-modality feature is present then it may be indicative of a disk/wind precession. \citep{2018ApJ...854..176V,2018ApJ...853..115W}.
Other possibilities discussed in the literature include; the propeller effect which requires the neutron star to have a very strong magnetic field \citep{1975A&A....39..185I,2016MNRAS.457.1101T}; binary systems in which the neutron star is typically in a moderately eccentric orbit (e < 0.5) \citep{1986ApJ...308..669S}, or the change in the accretion rate via flow spherization  as the system transition from the high to the low state \citep{2017AstL...43..464G}. In the framework of supercritical accretion, the wind emanating from the supercritical disk forms a funnel, which becomes visible during the precessional phases, thus leading to the visibility of a bright and spectrally hard X-ray source. In the other phases when the wind overlaps the channel (the angle at which the disk and the wind are seen exceeds the opening angle of the channel), the luminosity will decrease by orders of magnitude. SS433 \citep{2004ASPRv..12....1F,2007MNRAS.377.1187P,2018ApJ...853..115W} is an example of such a process.

Given the relative hard spectra and the long-term flux variability indicated particularly by the {\it Chandra} data set (see black points; Fig.\ref{F:3}), it is possible that ULX-4 may in fact host a NS \citep{2018ApJ...854..176V,2020ApJ...895...60R}. Of course, we note that a pulsar-like signal has not been recorded. As noted by \citet{2017PASJ...69...33O}, the absence of such a signal may be due to a line of sight highly inclined with respect to the binary system axis. Entertaining the NS scenario a little further, we can follow \cite{2016MNRAS.457.1101T}, and use the flux-variability factor $V_{f}$ (determined earlier), to estimate the spin period (P) of the NS. We obtain $\sim$ 1.5 s using the relation $\Delta$L $\sim$ 170P$^\frac{2}{3}$ M$_{1.4}$$^\frac{1}{3}$ R$_{6}^{-1}$, where $M_{1.4}$ is the NS mass in units of 1.4 $M\odot$, $R_{6}$ is NS radius in units of 10$^{6}$ cm. This spin period is comparable to some known PULXs (e.g. X-2 in M82, X-1 in NGC 5907 and X-2 in NGC 1313) \citep{2020MNRAS.495L.139T}.

Focusing more on the long-term spectral evolution of ULX-4, we note the recent study of \cite{2021A&A...649A.104G}, in which they presented a hardness-luminosity diagram (HLD) for a sample of 17 ULXs (including PULXs) with archival data taken from {\it Chandra}, {\it XMM-Newton} and {\it NuSTAR}. This diagram is equivalent to the hardness-intensity diagram (HID) \citep{2006ARA&A..44...49R,2010LNP...794...53B} that has been used for many years to probe the spectral evolution of galactic BH (and NS) binary systems and the internal coupling of the accretion process involving the corona and the accretion disk. A particular feature of the HID is the so-called q-curve shape that is traced out by BH transients as they undergo spectral transitions from the hard state through intermediate hard and soft states, to the  soft state, and finally back down to the hard state. The primary aim of the HID, and other diagrams such as the color-color diagram (CCD) introduced by \cite{2003MNRAS.342.1041D}, is to map the spectral transitions and, if possible, identify the nature of the compact object from the tracks they follow in the respective domains. In order to ascertain the position of ULX-4 on the HLD (HID), we imported the figure shown by {\it upper panel} of \cite{2021A&A...649A.104G} (their Figure 4) onto which we plotted the hardness-luminosity values we extracted for ULX-4; the result is displayed in the Fig.\ref{F:LXhard}. The {\it XMM-Newton} data are indicated as black stars and the {\it Chandra} set are denoted by red stars. The hardness was determined as the ratio of the unabsorbed fluxes in the hard (1.5 – 10 keV) to soft (0.3 – 1.5 keV) bands. As is seen in the {\it upper panel} of Fig.\ref{F:LXhard}, ULX-4 occupies the lower section of the luminosity scale and the middle region of the hardness scale. However, according to the findings of \cite{2021A&A...649A.104G}, the majority of the confirmed PULXs reside in the region corresponding to higher luminosity and higher hardness. This would tend to suggest that the compact object in ULX-4 is either a stellar BH or a NS with a low magnetic field. Moreover, we note that the {\it XMM-Newton} points essentially cluster around an average hardness ratio of $\sim 1.7$ and exhibit no discernible pattern. This is consistent with our discussion regarding the possible contribution of diffuse emission and unresolved background sources in the {\it XMM-Newton} spectra. In contrast, the {\it Chandra} points do show a hint of spectral evolution, with the hardness varying in the range $\sim (1.1 - 4.1)$, and trace out what appears to be a partial q-curve-like track. Taken at face value this would argue for an ordinary stellar BH as the compact object in ULX-4. Clearly, the q-curve track is not complete and one would need additional data to draw a definite conclusion. In a similar fashion, we further probe the nature of ULX-4 by deploying the CCD; in this case we define a hard color as the ratio of fluxes in the bands (5 - 8) and (2 - 5) keV, and a soft color as the ratio of fluxes in the bands (1 - 2) and (0.3 - 1) keV. Shown in the {\it bottom panel} of Fig.\ref{F:LXhard} is the resulting plot of hard color vs. soft color. We make a couple of observations regarding this result; the plot shows a diagonal track that is consistent with the one obtained by \cite{2003MNRAS.342.1041D} for a sample of galactic BH transients whose spectra are well described by a {\it power-law}. Secondly, they point out that the lower-left region of the plot (see their Figure 8) is only accessible to BH transients and not NS binaries (owing to the emission from the boundary layer between the NS surface and the inner part of the accretion disk). The ULX-4 points occupy precisely the noted region. This feature taken by itself would suggest that ULX-4 is more likely to be a transient BH than a NS.\\  

\subsection{Optical}

We have identified two possible optical counterparts of ULX-4 using the {\it HST}/ACS and {\it HST}/WFC3 data. The two counterparts were detected within the astrometric error radius of 0\farcs18. Their apparent magnitudes (m$_{V}$) are between 23 - 23.5 mag and their absolute magnitude (M$_{V}$) is $\sim$ $-$6.6 mag. These magnitudes are well in the range of other known ULXs \citep{2011ApJ...737...81T,2013ApJS..206...14G,2018ApJ...854..176V}.

\begin{figure}
\begin{center}
\includegraphics[width=\columnwidth]{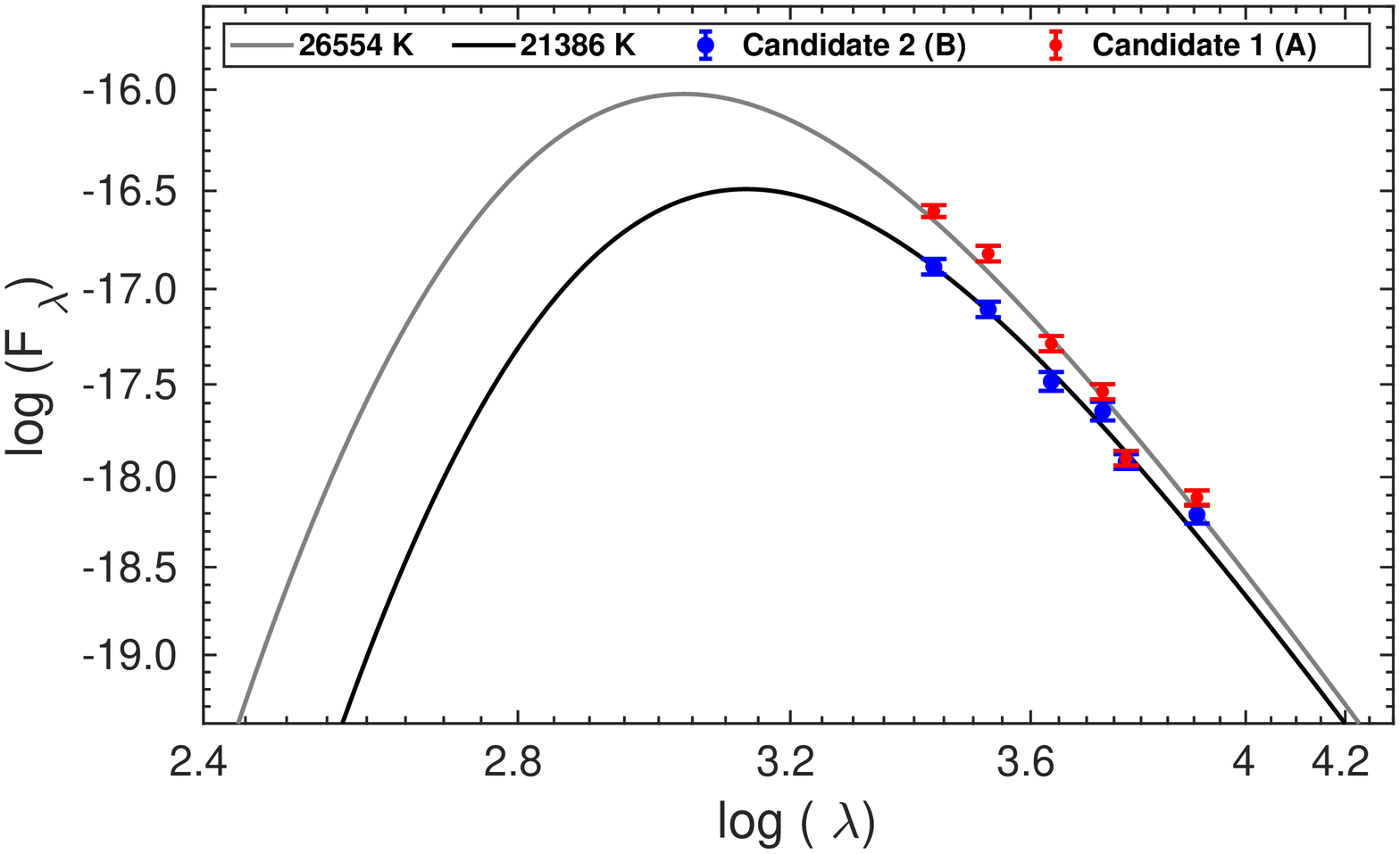}
\caption{The reddening corrected SEDs of optical counterparts A (red circles) and B (blue circles). The solid lines indicate blackbody models fitted to the available wavelength points, with temperatures of 26554 K for A and 21386 K for B. The units of y and x axes are erg s$^{-1}$ cm$^{-2}$ \AA$^{-1}$ and \AA, respectively.}
\label{F:BB}
\end{center}
\end{figure}

The spectral types of the counterparts A and B were estimated in two different ways: (1) Use of the CK04 models indicate both counterparts to be early B-type supergiants. As part of the same procedure, we identified the spectral types of two nearby field stars R1 and R2; these also turned out to be early B-type supergiants. (2) We made use of spectral fits using a blackbody function and obtained good fits to the SEDs and extracted temperatures of 26554 $\pm$ 104 K and 21386 $\pm$ 173 K for counterparts A and B respectively, with 95\% confidence level. The fits for the counterparts and the field stars are shown in Fig.\ref{F:9} and Fig.\ref{F:BB}. The B-type nature of the counterparts were identified using the temperature and luminosity-class tables in \cite{1981Ap&SS..80..353S}. 
Considering the moderate periodic modulation of source A, we also checked the spectral type of the donor using data taken simultaneously with filters ACS/WFC/F435W, F555W and F814W. The approximation by the models of stellar atmospheres with these three points confirms that a shape of the energy distribution corresponds to a B type supergiant.
The age and mass of the counterparts were determined as $\sim$ 10 Myr and $\sim$ 20 M$_{\odot}$, respectively; the corresponding CMD is shown in Fig.\ref{F:VVI}. The B-type classification for the counterparts of ULX-4 is consistent with the general finding of OB-type giants as optical counterparts for other ULXs.

Interestingly, the classification of the field sources R1 and R2 also turns out to be the same as that for the counterparts A and B, suggesting the possibility that these sources are part of the same star cluster cataloged by \cite{2016ApJ...824...71C}. As seen from Fig. \ref{F:6}, the optical counterparts are located within 18 pc to a nearby cluster (where 1$\arcsec$ corresponds to $\sim$ 44 pc for the adopted distance of M51). We estimate the age of the cluster as 6.7 Myr and a mass as $\sim$ 3.2$\times10^{4}$ M$_{\odot}$. \cite{2016ApJ...824...71C} find $\sim$ 6.8 Myr and 2.2$\times10^{4}$ M$_{\odot}$ as the age and the mass respectively. There is excellent agreement on the age but the mass we extract is a little higher than given by \cite{2016ApJ...824...71C}. We suspect the small difference in the derived mass is likely due to fact that we used a larger aperture size, 8 pixels, in our photometry compared to the 2.5 pixels used by \cite{2016ApJ...824...71C}.

One of our more intriguing results is that the optical counterpart A appears to exhibit a low-amplitude periodic modulation of 264$\pm$37 days although it appears only in the F606W filter. We consider two possibilities for this apparent periodicity: one involves the nature of the donor and the other about the nature of the periodicity itself. In the first case, the optical modulation could result from the orbital motion of the donor which is distorted by the effects of gravity and is not irradiated equally by the X-ray radiation \citep{2012MNRAS.419.1331Z}. Unfortunately, in this case the donor star can not be a B-type supergiant, since such a star is not likely to fill its Roche lobe assuming the periodicity is a result of the orbital period of several hundred days. Moreover, this implies the observed SED would have to be interpreted as a superposition of radiation from a hot wind, perhaps emanating from an accretion disk, and emission from a relatively cold donor. Such a donor may be a red supergiant (RSG) as has been noted in some ULX systems \citep{2019ApJ...883L..34H, 2020MNRAS.497..917L}. In particular, by assuming Roche-lobe overflow, \cite{2019ApJ...883L..34H} indicate that a red supergiant in NGC300 ULX-1 system is consistent with an orbital period in the range 0.8 - 2.1 years. This range accommodates the apparent periodicity we have found for ULX-4. 

In the second scenario, we assume the donor is indeed a B-type (in fact a Be/X) star therefore there is no need for reassessing the nature of the SED but it calls into question the nature of the underlying cause for the observed periodicity. Here we note the results reported by \cite{2020MNRAS.495L.139T}, where they display a reasonably tight positive correlation between the orbital period and the super-orbital period for a number of disk-fed high-mass X-ray binaries, Be/X binaries, and ULX pulsars. The super-orbital period is presumed to arise as the compact object interacts (regularly as part of its orbital motion) with a circumstellar disk of the donor and causes disturbances and/or obscurations in the medium of the disk which in turn propagate as density waves throughout the circumstellar disk causing a periodic feature in the observed optical light curves. This picture assumes that the periodicity we are witnessing is the super-orbital motion rather the ordinary orbit motion of the compact object. By assuming this scenario (i.e., a Be/X star since this class represents almost 60$\%$ of the high-mass X-ray binary population) and using the relation given by \cite{2020MNRAS.495L.139T}, one can estimate the orbital period for the compact object to be $\sim 10 - 15 $ days. An interesting consequence of this scenario is that the compact object is highly likely to be a NS since we know that from over a 100 known galactic Be/X binaries only one has been shown to host a BH (see \cite{2018MNRAS.477.4810B}, and \cite{2009ApJ...707..870B}). Moreover, if the compact object is indeed a NS then it is most likely to be in the super-Eddington accretion regime. Of course, we recognize that both of the scenarios outlined above fall short of a definitive conclusion, a situation not entirely surprising given the limited quality data available for the analysis.

\section{Summary and Conclusions}
\label{sec:4}

We have analyzed almost all the available X-ray data for the transient source, ULX-4, in M51. The data were taken from the {\it Chandra}, {\it XMM-Newton}, and {\it Swift-XRT} archives. In addition, we utilized optical data from the {\it HST} archives. The two main goals of the analysis were to a) determine the nature of ULX-4, and b) identify the optical counterpart(s) of the X-ray source. We performed relative astrometry between {\it Chandra} and {\it HST} data and identified two possible optical counterparts for ULX-4. By making use of FFTs and the construction of LS periodograms, both short-term and long-term variability searches, including the extraction of spectral lags for the X-ray light curves, were carried out for both the X-ray and the optical data sets. X-ray spectra were extracted and limited spectral analysis was done by deploying simple spectral models such as the {\it power-law} in order to ascertain and describe the main spectral features of the data. Additional spectral analysis of the X-ray data included the extraction of hardness ratios and the construction of color-color diagrams to identify possible spectral transitions of the source and to delineate the nature of the compact object in ULX-4. For the optical data, we constructed SEDs which were then fitted with a blackbody model to constrain the temperatures and obtain spectral types for the optical counterparts. CMDs were created to obtain the mass and age of the optical counterparts. Our main findings from this study are summarized as follows:
\begin{itemize}
\item We have identified two optical counterparts (labelled as A and B) for ULX-4; the age and mass of A is 9 Myr and 20 M$\sun$ respectively; that of B is 10 Myr and 18 M$\sun$ respectively, 
\item Based on the optical SEDs, the temperatures for the counterparts are (26554 $\pm$ 104 ) K for A and (21386 $\pm$ 173) K for B, suggesting a spectral type in the range B0I and B3I (but see note about optical variability down below),
\item Counterpart A exhibits a low-amplitude optical periodicity of 264 $\pm$ 37 days in the F606W filter; no periodicity was found for counterpart B, 
\item If we assume the apparent periodicity seen in A is associated with the orbital motion of the donor then the donor is more likely a red RSG in order to satisfy the long periodicity and the need to accommodate Roche-lobe overflow. As a consequence, the SED would then have to be interpreted as a superposition of emissions from a cold donor and a hot flow component most likely from an accretion disk. If, on the other hand, the periodicity is super-orbital in nature i.e., due to interactions of the compact object with a circumstellar disk, then the donor could be a Be/X star with the compact object being a NS, 

\item The majority of the X-ray data are best described by a simple {\it power-law} with an index in the range 1.44 - 2.0; a couple of {\it XMM-Newton} observations indicate the presence of a soft component which are best described with the addition of a {\it mekal} component suggesting a contribution from a diffuse background source, 
\item We find some evidence for the existence of a bi-modal distribution of the X-ray flux of ULX-4, the cause of which is of some debate in the literature \citep{1975A&A....39..185I,1986ApJ...308..669S,2018ApJ...854..176V}
\item The HID, the CCD, along with the significant variability noted in the long-term X-ray light curve from {\it Chandra}, present a rather mixed picture as to the nature of ULX-4 i.e., arguments could be advanced for either a stellar-size BH or a low-magnetic field NS. Unfortunately, the data are incomplete and/or of low statistics thus rendering a definitive conclusion speculative at best.
\item Finally, as a side product of this study, we note the presence of a previously uncatalogued (transit) X-ray source (CXOU J132951.7 +471010) in a number of {\it Chandra} observations. The unabsorbed flux of this new source in the range of (0.05$-$1.88)$\times$10$^{-14}$ erg cm$^{-2}$ s$^{-1}$.
\end{itemize}

\begin{table*}
\centering
\caption{ X-ray observations log for ULX-4 in M51}
\begin{tabular}{ccccccccc}
\hline\hline
 Observatory & Label & ObsID & Instrument & Date & Exp. \\
& & & & (YYYY-MM-DD) & (ks) \\
\hline
{\it XMM-Newton}
& XM1 & 0212480801 & EPIC & 2005-07-01 & 49.21  \\
& XM2 & 0303420101 & EPIC & 2006-05-20 & 54.11  \\
& XM3 & 0303420201 & EPIC & 2006-05-24 & 36.81  \\
& XM4 & 0677980701 & EPIC & 2011-06-07 & 13.32  \\
& XM5 & 0677980801 & EPIC & 2011-06-11 & 13.32  \\	
& XM6 & 0830191501 & EPIC & 2018-06-13 & 63.00  \\
& XM7 & 0830191601 & EPIC & 2018-06-15 & 63.00  \\
& XM8 & 0852030101 & EPIC & 2019-07-11 & 77.00  \\
\hline
{\it Chandra}	
& C1    & 354      & ACIS-S &2000-06-20 & 14.86 \\
& C2    & 1622     & ACIS-S & 2001-06-23 & 26.81 \\
& C3    & 3932     & ACIS-S & 2003-08-07 & 47.97 \\
& C4    & 12562    & ACIS-S & 2011-06-12 & 9.63 \\
& C5    & 12668    & ACIS-S & 2011-07-03 & 9.99   \\
& C6    & 13813    & ACIS-S & 2012-09-09 & 179.20 \\
& C7    & 13812    & ACIS-S & 2012-09-12 & 157.46 \\
& C8    & 15496    & ACIS-S & 2012-09-19 & 40.97 \\
& C9    & 13814    & ACIS-S & 2012-09-20 & 189.85 \\
& C10   & 13815    & ACIS-S & 2012-09-23 & 67.18 \\
& C11   & 13816    & ACIS-S & 2012-09-26 & 73.10 \\
& C12   & 15553    & ACIS-S & 2012-10-10 & 37.57 \\
& C13   & 19522    & ACIS-I	& 2017-03-17 & 37.76 \\
& C14   & 20998    & ACIS-S & 2018-08-31 & 19.82 \\
\hline
\end{tabular}
\label{T:1}
\end{table*}

\begin{table*}
\centering
\begin{minipage}[b]{0.9\linewidth}
\caption{X-Ray spectral fitting parameters of ULX-4.}
\begin{tabular}{ccccccccccc}
\hline
\hline
Label & Count rate & N$_{\mathrm{PL}}$ & N$_{\mathrm{mekal}}$ & ${\Gamma}$ & kT & F$_{\mathrm{X}}$ & L$_{\mathrm{X}}$ & $\chi^{2}$/dof ($\chi^{2}_V$)&\\
(1) & (2) & (3) & (4) & (5) & (6) & (7) & (8) & (9)\\
\hline
\multicolumn{10}{c}{Model: {\it tbabs} $\times$ {\it power-law}}  \\
\hline
C1 & <0.11 & ... & ... & ...& ... & <0.04  & <0.04 &... \\
C2 & 23.87 $\pm$ 0.96 & $2.40_{-0.19}^{+0.19}$ & ... & $1.44_{-0.11}^{+0.11}$ & ... & $2.12_{-0.15}^{+0.15}$  & $2.05_{-0.14}^{+0.14}$ & 54.51/53(1.03) \\
C3 & <0.06  & ... & ... & ...& ... & <0.02  & <0.02 &... \\
C4 & 10.59 $\pm$ 1.09 & $1.52_{-0.34}^{+0.36}$ & ... & $1.95_{-0.32}^{+0.32}$ & ... & $0.88_{-0.17}^{+0.17}$  & $0.85_{-0.16}^{+0.16}$ & 15.76/15(1.05) \\
C5 & 0.35 $\pm$ 0.02  & ... & ... & ...& ... & 0.05 $\pm$ 0.02  & 0.05 $\pm$ 0.02 &... \\
C6 & 0.13 $\pm$ 0.06 & ... & ... & ...& ... & 0.014  $\pm$ 0.007  & 0.02  $\pm$ 0.01 &... \\
C7 & 0.10 $\pm$ 0.04 & ... & ... & ...& ... & 0.011  $\pm$ 0.006  & 0.01  $\pm$ 0.01 &... \\
C8 & 0.07 $\pm$ 0.02  & ... & ... & ...& ... & 0.007  $\pm$ 0.0004  & 0.01  $\pm$ 0.01 &... \\
C9 & 0.14 $\pm$ 0.08 & ... & ... & ...& ... & 0.02  $\pm$ 0.01  & 0.02  $\pm$ 0.01 &... \\
C10 & 0.15 $\pm$ 0.02 & ... & ... & ...& ... & 0.02  $\pm$ 0.01  & 0.02  $\pm$ 0.01 &... \\
C11 & 3.24 $\pm$ 0.23 & $0.44_{-0.07}^{+0.07}$ & ... & $1.63_{-0.22}^{+0.23}$ & ... & $0.32_{-0.04}^{+0.04}$  & $0.31_{-0.04}^{+0.04}$ & 34.65/43(0.81) \\
C12 & 22.41 $\pm$ 0.79 & $3.45_{-0.25}^{+0.25}$ & ... & $1.75_{-0.10}^{+0.10}$ & ... & $2.29_{-0.14}^{+0.14}$  & $2.21_{-0.13}^{+0.13}$ & 63.77/66(0.96) \\
C13 & 0.13 $\pm$ 0.02 & ... & ... & ...& ... & 0.03 $\pm$ 0.01 & 0.03  $\pm$ 0.01 &... \\
C14 & 8.58 $\pm$ 0.69 & $1.81_{-0.42}^{+0.44}$ & ... & $1.78_{-0.29}^{+0.30}$ & ... & $1.16_{-0.17}^{+0.17}$  & $1.13_{-0.16}^{+0.16}$ & 31.41/28(1.12) \\
XM1 & 39.20 $\pm$ 1.51 & $2.52_{-0.15}^{+0.15}$ & ... & $1.85_{-0.11}^{+0.11}$ & ... & $1.55_{-0.09}^{+0.09}$  & $1.50_{-0.09}^{+0.09}$ & 126.63/83(1.53) \\
XM2 & 21.60 $\pm$ 1.45 & $1.36_{-0.13}^{+0.13}$ & ... & $2.00_{-0.19}^{+0.19}$ & ... & $0.76_{-0.07}^{+0.07}$  & $0.74_{-0.07}^{+0.07}$ & 59.71/69(0.87) \\
XM3$^{*}$ & 9.21 $\pm$ 1.10 & ... & ... & ...& ... & 0.25 $\pm$ 0.03  & 0.24 $\pm$ 0.03 &... \\
XM4 & 50.60 $\pm$ 2.61 & $3.14_{-0.29}^{+0.28}$ & ... & $1.90_{-0.17}^{+0.17}$ & ... & $1.87_{-0.12}^{+0.12}$  & $1.81_{-0.11}^{+0.11}$ & 23.64/15(1.58) \\
XM5 & 31.90 $\pm$ 4.75 & $1.84_{-0.38}^{+0.38}$ & ... & $1.73_{-0.47}^{+0.47}$ & ... & $1.23_{-0.26}^{+0.26}$  & $1.19_{-0.25}^{+0.25}$ & 24.26/29(0.84) \\
XM6 & 18.90 $\pm$ 0.80 & $1.27_{-0.09}^{+0.09}$ & ... & $1.86_{-0.12}^{+0.12}$ & ... & $0.78_{-0.06}^{+0.06}$  & $0.75_{-0.05}^{+0.05}$ & 51.94/38(1.37) \\
XM7 & 17.40 $\pm$ 0.78 & $1.07_{-0.08}^{+0.09}$ &...& $1.84_{-0.14}^{+0.14}$ & ... & $0.60_{-0.05}^{+0.05}$  & $0.59_{-0.05}^{+0.05}$ & 19.17/20(0.96) \\
XM8$^{*}$ & 9.52 $\pm$ 0.63 & ... & ... & ...& ... & 0.26 $\pm$ 0.02 & 0.25 $\pm$ 0.02 &... \\
\hline
\multicolumn{10}{c}{ Model: {\it tbabs} $\times$ ({\it power-law} + {\it mekal}) } \\
\hline
XM1 & ... & $1.99_{-0.15}^{+0.15}$ & $5.17_{-1.06}^{+1.06}$ & $1.68_{-0.13}^{+0.13}$ & $0.59_{-0.17}^{+0.17}$ & $1.55_{-0.05}^{+0.05}$  & $1.50_{-0.09}^{+0.09}$ & 99.37/81(1.23) \\
XM4 & ... & $2.43_{-0.56}^{+0.56}$ & $8.38_{-4.89}^{+4.89}$ & $1.57_{-0.32}^{+0.32}$ & $0.30_{-0.08}^{+0.26}$ & $2.08_{-0.08}^{+0.08}$ & $2.02_{-0.08}^{+0.08}$ & 18.05/13(1.39) \\
XM5 & ... & $1.18_{-0.33}^{+0.33}$ & $7.63_{-3.77}^{+3.77}$ & $1.10_{-0.29}^{+0.29}$ & $0.25_{-0.10}^{+0.10}$ & $1.83_{-0.12}^{+0.12}$ & $1.77_{-0.10}^{+0.11}$ & 21.78/27(0.80) \\
\hline
\end{tabular}
\\ Note. — Col. (1): Observation label corresponding to Column (1) of Table \ref{T:1}. Col. (2): Source count rates in units of $10^{-3}$ count$s^{-1}$ in 0.3$-$10 keV. Col. (3): N$_{\mathrm{PL}}$ is the normalization of the {\it power-law} model at 1 keV in units of $10^{-5}$ photons $cm^{-2}$ s$^{-1}$. Col. (4): N$_{\mathrm{mekal}}$ is the normalization of the {\it mekal} model in units of $10^{-6}\times10^{-14}/4\pi(D(1+z))^{2}\int n_{e}n_{H}dV$ where D is the angular diameter distance to the source (cm), $n{_e}$ and $n{_H}$ are the electron and H densities (cm$^{-3}$), respectively. Col. (5): $\Gamma$ is the photon index of the {\it power-law} model. Col. (6): Plasma temperature in unit of keV. Col. (7): Unabsorbed flux in units of 10$^{-13}$ ergs $cm^{-2}$ $s^{-1}$.
Col. (8): Unabsorbed luminosity in units of 10$^{39}$ ergs $s^{-1}$. Col. (9): Best-fit $\chi^{2}$ and degrees of freedom. The $\chi^2_{\nu}$ is given in parentheses. All errors are at 90\% confidence level. Unabsorbed flux and luminosity values were calculated in the 0.3–10 keV energy band. Adopted distance of 9 Mpc \citep{2014AJ....148..107R,2020MNRAS.491.1260S} was used for luminosity. $^{*}$Since both XM3 and XM8 have insufficient data the source fluxes were calculated using the Webpimms tool with a photon index of $\Gamma$ =1.75 and N$_{H}$=0.03$\times$10$^{22}$ cm$^{2}$. Except for XM1, simultaneous fit of EPIC pn + MOS spectra for the remaining datasets did not give us acceptable parameters for a single and two component models. Only the X-ray absorption value N$_{H}$=0.03$\times$10$^{22}$ cm$^{2}$ as used for all spectra and this value was kept fixed.\\
\label{T:2}
\end{minipage}
\end{table*}

\begin{table}
  \caption{The log of {\it HST}/ACS and WFC3 observations.}
  \label{tab:freq}
  \begin{tabular}{cccccl}
\hline\hline
 Instrument & 
Filter & ObsID & Date & Exp. \\
& & & (YYYY-MM-DD) & (s)\\
\hline
ACS/WFC & F435W & J97C51R3Q & 2005-01-20  & 340 \\
ACS/WFC & F555W & J97C51R4Q & 2005-01-20  & 340 \\
ACS/WFC & F814W & J97C51R6Q & 2005-01-20  & 340 \\
\hline
ACS/WFC3 & F275W & ICD401010 & 2014-09-11 & 1400 \\
ACS/FWC3 & F336W & ICD401020 & 2014-09-11 & 1450 \\
\hline
ACS/WFC & F606W & JD8F01010 & 2016-10-05 & 2200 \\
\hline
 \label{T:3}
\end{tabular}
\end{table}

\begin{table*}
\centering
\caption{Coordinates of the X-ray/optical reference sources.}
\begin{tabular}{ccccccccccc}
\hline
\hline
Source. & {\it Chandra} R.A.& {\it Chandra} Dec.& Counts$^{a}$ & {\it HST} R.A.& {\it HST} Dec. &  \\
& (hh:mm:ss.sss) & ($\degr$ : $\arcmin$ : $\arcsec$) & - & (hh:mm:ss.sss) & ($\degr$ : $\arcmin$ : $\arcsec$) \\
\hline
\multicolumn{6}{c}{{\it Chandra} ACIS-S X-ray sources (ObsID 13816) identified in {\it HST} observation (J97C51R4Q)} &Offset(${\arcsec}$) \\
\\
SN 2011dh & 13:30:05.108 & +47:10:11.10 & 79.25 ${\pm 10.63}$ & 13:30:05.104 & +47:10:10.89 & 0.21\\
XOUJ133006.5+470834 & 13:30:06.456 & +47:08:34.86 & 348.71${\pm 21.74}$ & 13:30:06.446 & +47:08:34.72 & 0.17\\
J133011+471041 & 13:30:11.023 & +47:10:41.24 & 143.93 ${\pm 14.57}$& 13:30:11.009 & +47:10:41.14 & 0.18\\
\\
\multicolumn{6}{c}{{\it Chandra} and corrected optical coordinates of optical counterparts} & Position uncertainty (${\arcsec}$)\\
\\
ULX-4 & 13:29:53.333 & +47:10:42.76 & ... & 13:29:53.323 & +47:10:42.64 & 0.18\\
\hline
\end{tabular}
\\Note. — $^{a}$: The {\it Chandra} counts were calculated in the 0.3-10 keV using {\scshape xspec} \\
\label{T:4}
\end{table*}

\begin{table*}
\centering
\caption{The dereddened Vega magnitudes and flux values of optical counterparts of ULX-4}
\begin{tabular}{cccccccccc}
\hline\hline
Filter & Pivot Wavelength$^{a}$ & ZP$^{b}$ & Aperture corr. & \multicolumn{2}{c} {Vega Magnitude} & \multicolumn{2}{c} {Flux}\\
& (\AA) & (mag) & (mag) &&&\multicolumn{2}{c}({$10^{-14}$ erg $s^{-1}$ $cm^{-2}$})\\
\hline
& & & & A & B & A & B \\
\hline
F275W  & 2710.4 & 22.66 & 0.43 & $21.25\pm 0.03$ & $21.96\pm 0.04$ &  6.78 $\pm$ 0.04 & 3.53 $\pm$ 0.05\\
F336W  & 3354.8 & 23.46 & 0.38 & $21.45\pm 0.04$ & $22.17\pm 0.04$ &  5.10 $\pm$ 0.04 & 2.63 $\pm$ 0.05\\
F435W  & 4329.8 & 25.77 & 0.35 & $23.17\pm 0.04$ & $23.67\pm 0.05$ &  2.25 $\pm$ 0.04 & 1.42 $\pm$ 0.05\\
F555W  & 5360.8 & 25.71 & 0.27 & $23.12\pm 0.04$ & $23.40\pm 0.05$ &  1.55 $\pm$  0.04 &1.22 $\pm$ 0.05\\
F606W  & 5921.0 & 26.41 & 0.26 & $23.17\pm 0.04$ & $23.67\pm 0.04$ &  0.75 $\pm$ 0.04 & 0.72 $\pm$ 0.05\\
F814W  & 8044.8 & 25.52 & 0.28 & $23.19\pm 0.04$ & $23.35\pm 0.05$ &  0.62 $\pm$ 0.04 & 0.50 $\pm$ 0.05\\ 
\hline
\end{tabular}
\\Note. — $^{a}$ pivot wavelengths and $^{b}$ Zeropoint Vega magnitude of filters were calculated with {\scshape pysynphot}.\\
\label{T:5}
\end{table*}

{\begin{figure}
\begin{center}
\includegraphics[width=\columnwidth]{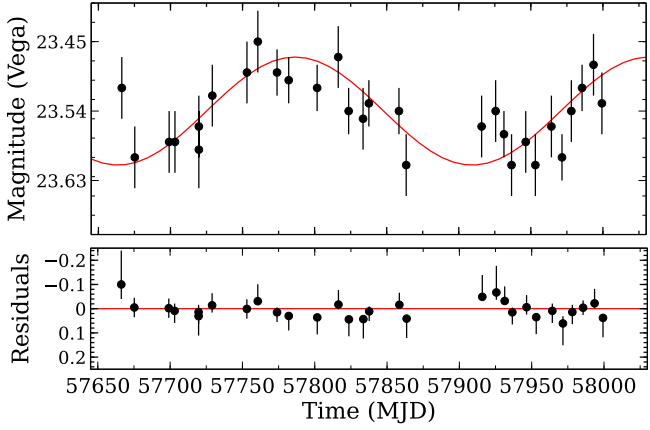}
\caption{Sine curve fit and residuals for the light curve of counterpart of A in the F606W observations. {\it HST} observations and sinusoidal fit represent filled black circles and solid red line, respectively. The residuals (data-model) are shown in the lower panel.}
\label{F:10}
\end{center}
\end{figure}

\begin{figure}
\begin{center}
\includegraphics[width=\columnwidth]{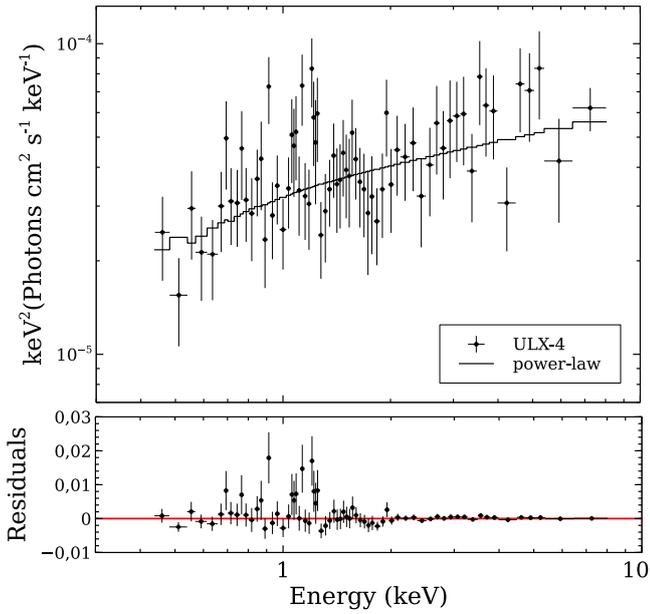}
\caption{The upper panel shows the spectra unfolded through the {\it power-law} model from C12 data. The residuals (data-model) are shown in the lower panel.}
\label{F:specraChandra}
\end{center}
\end{figure}

\begin{figure}
\begin{center}
\includegraphics[width=\columnwidth]{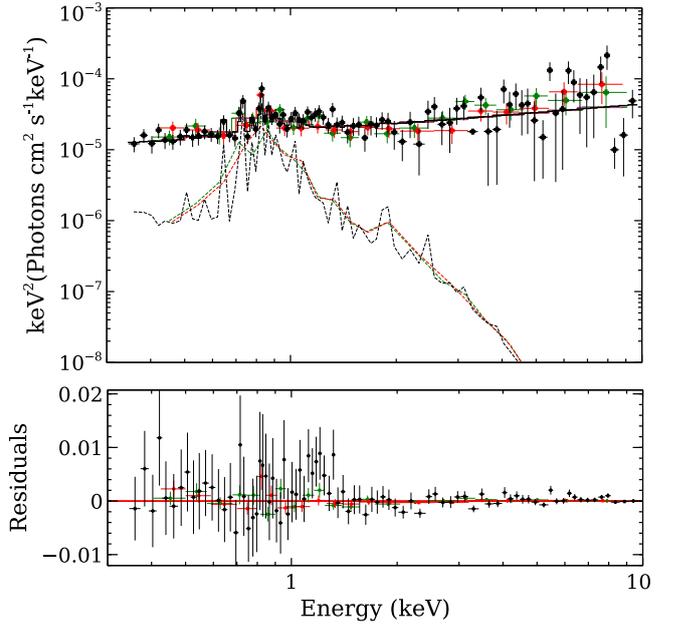}
\caption{The upper panel shows the spectra unfolded through the {\it power-law+mekal} model from XM1 data. The residuals (data-model) are shown in the lower panel. The dashed black, red, and green lines represent the {\it mekal} model for pn, MOS-1, and MOS-2 data, respectively.}
\label{F:XM1_uns}
\end{center}
\end{figure}

\begin{figure}
\begin{center}
\includegraphics[width=\columnwidth]{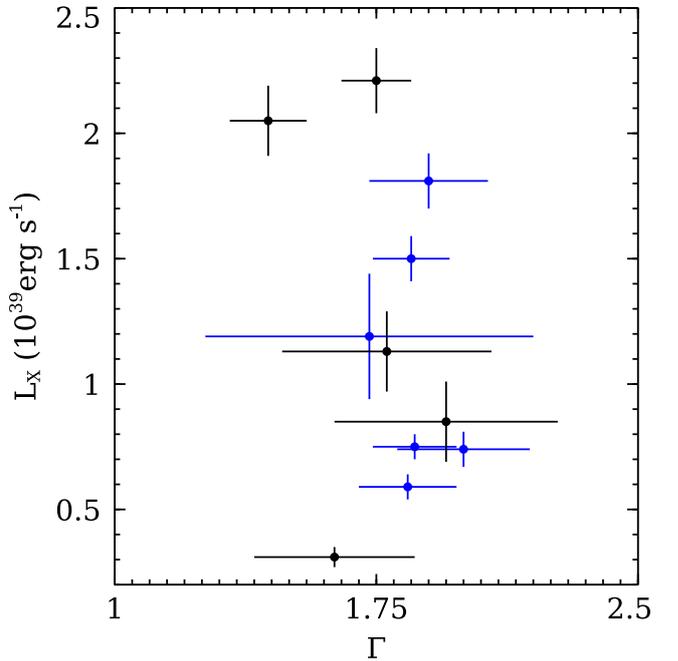}
\caption{ L$_{X}$ derived from {\it power-law} model vs. photon index of $\Gamma$ (see Table \ref{T:2}). Black and blue filled circles represent {\it Chandra} and {\it XMM-Newton} data, respectively.}
\label{F:GammaLX}
\end{center}
\end{figure}

\begin{figure}
\begin{center}
\includegraphics[width=\columnwidth]{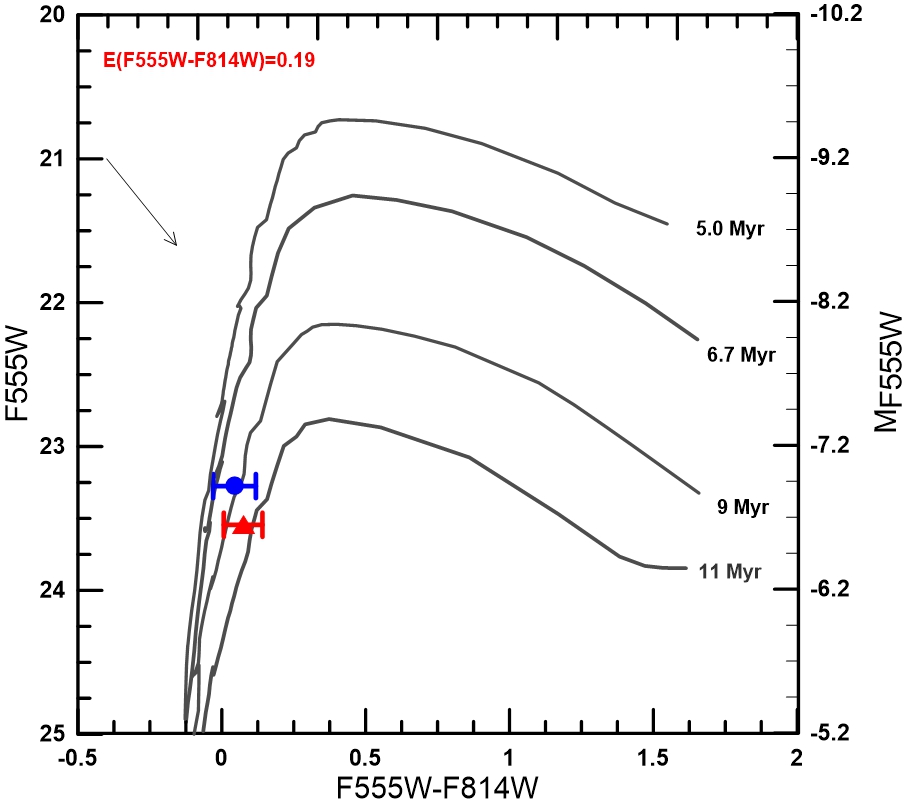}
\caption{The {\it HST}/ACS Color-Magnitude Diagram (CMD) for optical counterparts of ULX-4. Padova isochrones of different ages are overplotted. The red triangle and blue circle represent counterparts A and B, respectively. These isochrones have been corrected for extinction A$_{V}$ = 0.46 mag and the reddening line was shown black arrow.}
\label{F:VVI}
\end{center}
\end{figure}

\begin{figure}
\begin{center}
\includegraphics[width=\columnwidth]{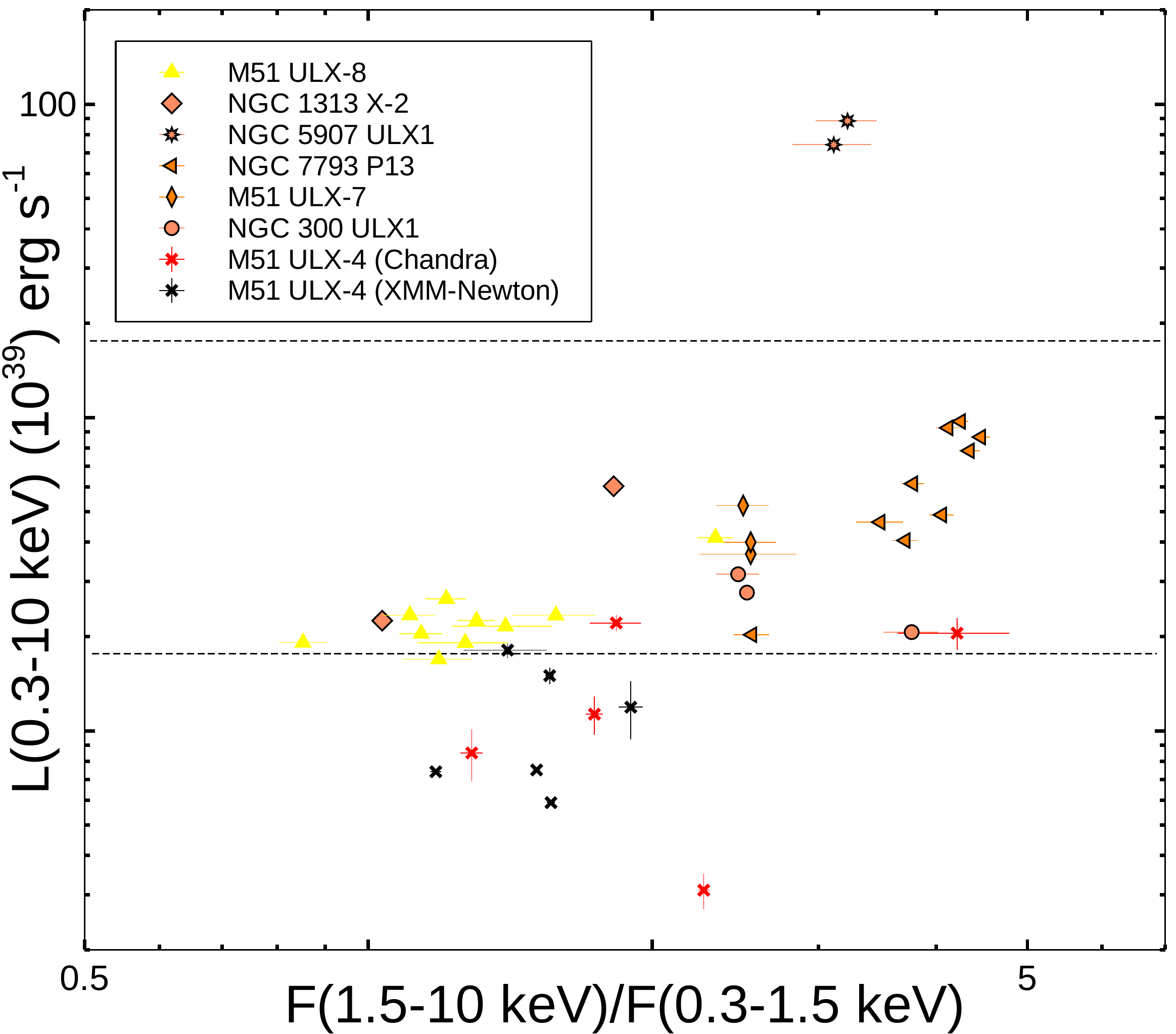}
\includegraphics[width=\columnwidth]{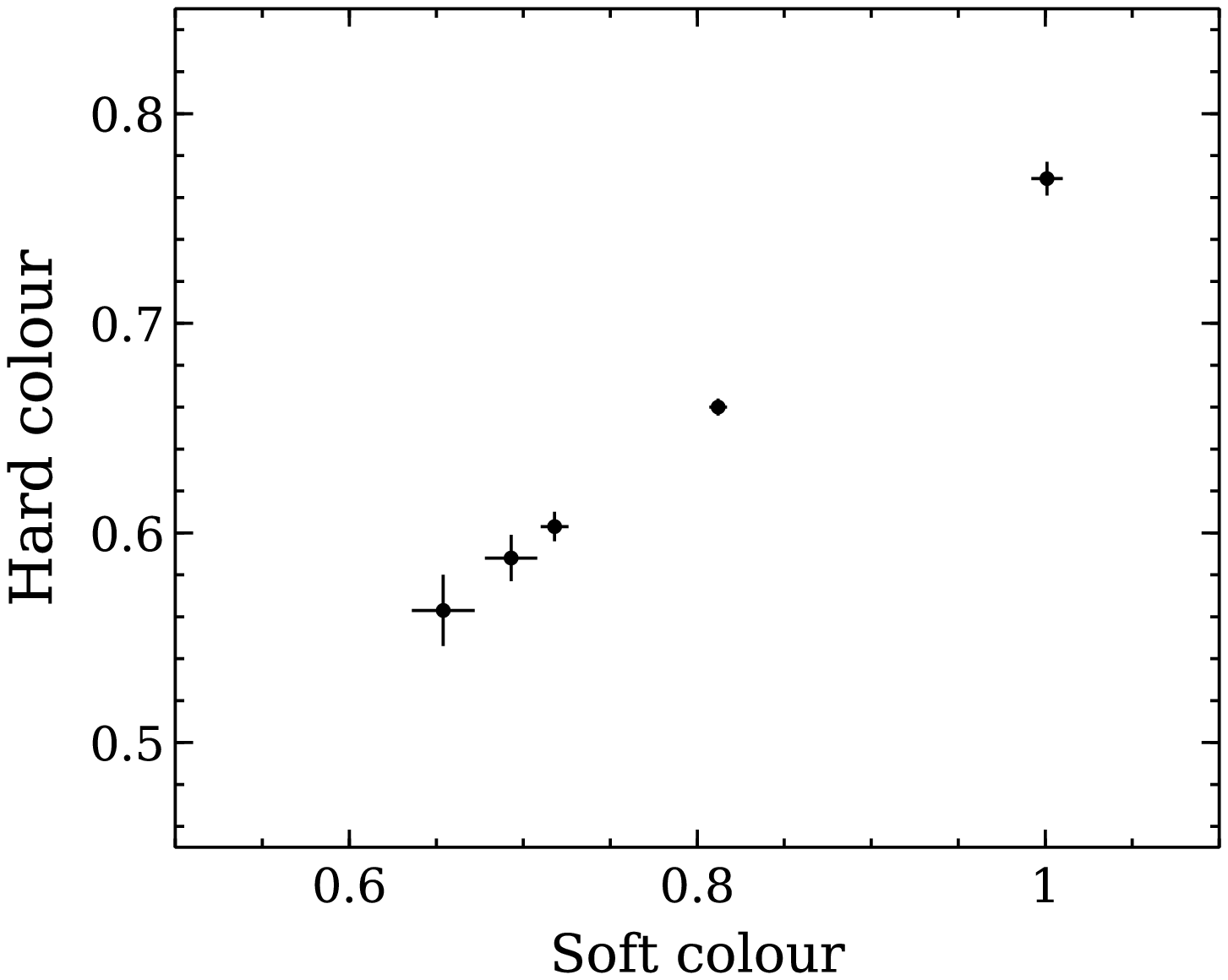}
\caption{The upper panel, we imported
the hardness-luminosity diagram of \protect\cite{2021A&A...649A.104G} (their Figure 4) onto which we plotted the hardness-luminosity values extracted for ULX-4; {\it Chandra} and {\it XMM-Newton} data are denoted by red and black stars, respectively. Yellow triangles and in shades of orange represent NS ULX and pulsating ULXs, respectively. The dashed black lines show 10 and 100 times the Eddington limit for a NS. The hardness is determined as the ratio of the unabsorbed fluxes in the hard (1.5 – 10 keV) to soft (0.3 – 1.5 keV) bands. The lower panel: Colour–colour diagram for the {\it Chandra} data. Hard colour is defined as the ratio of the fluxes in the 5–8 and 2–5 keV while soft colour is defined as the ratio of the 1–2 and 0.3-1 keV.}
\label{F:LXhard}
\end{center}
\end{figure}

\begin{table}
  \caption{The dereddened Vega magnitudes of the counterpart A obtained from ACS/WFC F606W observations.}
 \begin{tabular}{ccl}
\hline\hline
Date & Magnitude  \\
(MJD) & (Vega)\\
\hline
57666.23008	&	23.51	$\pm$	0.04	\\
57675.23803	&	23.60	$\pm$	0.04	\\
57699.07316	&	23.58	$\pm$	0.04	\\
57703.11242	&	23.58	$\pm$	0.04	\\
57719.68467	&	23.59	$\pm$	0.05	\\
57719.68467	&	23.56	$\pm$	0.04	\\
57729.02204	&	23.52	$\pm$	0.04	\\
57753.05545	&	23.49	$\pm$	0.04	\\
57760.60783	&	23.45	$\pm$	0.04	\\
57773.93019	&	23.49	$\pm$	0.03	\\
57782.01584	&	23.50	$\pm$	0.03	\\
57801.73972	&	23.51	$\pm$	0.03	\\
57816.38990	&	23.47	$\pm$	0.04	\\
57823.60626	&	23.54	$\pm$	0.03	\\
57833.60502	&	23.55	$\pm$	0.04	\\
57837.64349	&	23.53	$\pm$	0.03	\\
57858.50069	&	23.54	$\pm$	0.03	\\
57863.53201	&	23.61	$\pm$	0.04	\\
57915.85382	&	23.56	$\pm$	0.04	\\
57925.58697	&	23.54	$\pm$	0.04	\\
57931.34880	&	23.57	$\pm$	0.03	\\
57936.61826	&	23.61	$\pm$	0.04	\\
57946.45948	&	23.58	$\pm$	0.04	\\
57953.14623	&	23.61	$\pm$	0.04	\\
57964.26810	&	23.56	$\pm$	0.04	\\
57971.61629	&	23.60	$\pm$	0.03	\\
57978.10350	&	23.54	$\pm$	0.04	\\
57985.54818	&	23.51	$\pm$	0.03	\\
57993.48529	&	23.48	$\pm$	0.04	\\
57999.30881	&	23.53	$\pm$	0.04	\\
\hline
\label{T:6}
\end{tabular}
\end{table}

\section{ACKNOWLEDGEMENTS}
This research was supported by the Scientific and Technological Research Council of Turkey (TÜBİTAK) through project number 117F115. This research is a part of the PhD thesis of S. Allak and he acknowledges financial support by TÜBİTAK. ES and KD acknowledges support provided by the Scientific and Technological Research Council of Turkey (TÜBİTAK) through project number 119F334. We thank the TÜBİTAK National Observatory (TUG) for support with observing times and equipment. Astrometry was performed as part of the government contract of the SAO RAS approved by the Ministry of Science and Higher Education of the Russian Federation. The study of the nature of optical radiation was supported by the Russian Science Foundation (project no. 21-72-10167 ULXs: wind and donors)}. Finally, we acknowledge the useful comments and recommendations of the Referee which helped to clarify a number of issues.

\section*{Data Availability}
The scientific results reported in this article are based on archival observations made by the {\it Chandra}\footnote{https://cda.harvard.edu/chaser/}, {\it XMM-Newton}\footnote{http://nxsa.esac.esa.int/nxsa-web/} and {\it Swift-XRT}\footnote{https://www.swift.ac.uk/swift\_portal/} X-ray Observatories. This work has also made use of observations made with the NASA/ESA Hubble Space Telescope, and obtained from the data archive at the Space Telescope Science Institute\footnote{https://mast.stsci.edu/portal/Mashup/Clients/Mast/Portal.html}

\bibliographystyle{mnras}
\bibliography{m51} 
\bsp	
\label{lastpage}
\end{document}